\documentclass[11pt]{article}
\usepackage{cite}
\usepackage{amsmath,amsfonts,amssymb}
\usepackage[small,bf,hang]{caption}
\usepackage{slashed}
\usepackage{mathabx}
\usepackage{latexsym,epsfig}

\def\hybrid{
        \topmargin -20pt
        \oddsidemargin 0pt
        \headheight 0pt \headsep 0pt
        \textwidth 6.25in 
        \textheight 9.5in 
        \marginparwidth .875in
        \parskip 5pt plus 1pt \jot = 1.5ex}

\hybrid

 \csname
@addtoreset\endcsname{equation}{section}

\def\moth{\mathsurround=0pt}
\newdimen\zo \zo=0pt

\def\tick{\leaders\hrule height 0.5ex depth 0pt \hskip 0.5pt}
\def\upboxfill{$\moth \setbox\zo\hbox{\tick}%
  \hskip 3pt\hbox to 0pt{$\tick$\hss}\hrulefill \hbox to 7.5pt{$\tick$\hss}$}

\def\dtick{\leaders\hrule height .34pt depth 0.5ex \hskip 0.5pt}
\def\downboxfill{$\moth \setbox\zo\hbox{\dtick}%
  \hskip 2pt\hbox to 0pt{$\dtick$\hss}\hrulefill \hbox to 2pt{$\dtick$\hss}$}

\def\bec{\begin{center}}
\def\ec{\end{center}}

\def\be{\begin{equation}}
\def\ee{\end{equation}}
\def\bea{\begin{eqnarray}}
\def\eea{\end{eqnarray}}
\def\ba{\begin{array}}
\def\ea{\end{array}}

\begin{document}

\begin{titlepage}
\rightline{}
\rightline{\tt }
\rightline{\tt  MIT-CTP-4557}
\rightline{June 2014}
\begin{center}
\vskip 1.0cm
{\Large \bf {Exceptional Field Theory III:\, E$_{8(8)}$}}\\
\vskip 1.6cm
{\large {Olaf Hohm${}^1$ and Henning Samtleben${}^2$}}
\vskip .6cm
{\it {${}^1$Center for Theoretical Physics}}\\
{\it {Massachusetts Institute of Technology}}\\
{\it {Cambridge, MA 02139, USA}}\\
ohohm@mit.edu
\vskip 0.2cm
{\it {${}^2$Universit\'e de Lyon, Laboratoire de Physique, UMR 5672, CNRS}}\\
{\it {\'Ecole Normale Sup\'erieure de Lyon}}\\
{\it {46, all\'ee d'Italie, F-69364 Lyon cedex 07, France}}\\
henning.samtleben@ens-lyon.fr

\vskip 1.5cm
{\bf Abstract}
\end{center}

\vskip 0.2cm

\noindent
\begin{narrower}
We develop exceptional field theory for E$_{8(8)}$, 
defined on a (3+248)-dimensional generalized spacetime with extended coordinates 
in the adjoint representation of  E$_{8(8)}$. The fields transform under E$_{8(8)}$ generalized 
diffeomorphisms and are subject to covariant section constraints. 
The bosonic fields include an `internal' dreibein 
and an  
E$_{8(8)}$-valued `zweihundertachtundvierzigbein' (248-bein). 
Crucially, the theory also features  
gauge vectors for the E$_{8(8)}$ E-bracket governing the generalized diffeomorphism algebra  
and covariantly constrained gauge vectors for a separate but constrained E$_{8(8)}$ gauge symmetry. 
The complete bosonic theory, with a novel Chern-Simons term for the gauge vectors, 
is uniquely determined by gauge invariance under 
internal and external generalized diffeomorphisms. The theory consistently comprises 
components of the dual graviton encoded in the 248-bein. 
Upon picking particular solutions of the constraints the theory reduces to $D=11$ or type IIB supergravity, 
for which  
the dual graviton becomes pure gauge. This resolves the  
dual graviton problem, as we discuss in detail.

\end{narrower}

\end{titlepage}

\newpage

\section{Introduction}
In this paper we present the details of the recently announced `exceptional field theory' 
(EFT)~\cite{Hohm:2013pua} for the group E$_{8(8)}$, 
complementing the construction for E$_{6(6)}$ and E$_{7(7)}$ given in \cite{Hohm:2013vpa} and
\cite{Hohm:2013uia}, respectively.
The approach is a generalization of double field theory (DFT) 
\cite{Siegel:1993th,Hull:2009mi,Hull:2009zb,Hohm:2010jy,Hohm:2010pp,Hohm:2010xe},\footnote{See \cite{Hohm:2013bwa}
for a review and further references.} 
with the goal to render the dynamics of  the complete $D=11$ supergravity \cite{Cremmer:1978km}, 
and that of type IIB~\cite{Schwarz:1983wa,Howe:1983sra},  covariant
under the exceptional groups that are known to appear under
dimensional reduction~\cite{Cremmer:1979up}. 
We refer to the introduction of \cite{Hohm:2013vpa} for a more detailed outline of the general ideas,
previous approaches, and extensive references. Here we will mainly present and discuss 
the novel aspects relevant for the group  E$_{8(8)}$ which brings in some distinctive new features
as compared to the formulations for the smaller exceptional groups.

The E$_{8(8)}$ EFT is based on a generalized $3+248$ dimensional spacetime, with 
the `external' spacetime coordinates $x^{\mu}$ and `internal' coordinates $Y^M$
in the adjoint representation $\bf{248}$ of E$_{8(8)}$, 
with dual derivatives $\partial_M$.\footnote{Such 
generalized spacetimes also appear in the proposal of~\cite{West:2003fc}.}
The dependence of all fields on the extended 248 coordinates $Y^M$ is restricted 
by E$_{8(8)}$ covariant section constraints~\cite{Coimbra:2011ky,Berman:2012vc}
that project out sub-representations 
in the tensor product ${\bf 248}\otimes {\bf 248}$, 
\be\label{secConstrIntro}
  \big(\mathbb{P}_{1+248+3875}\big)_{MN}{}^{KL} \partial_K\otimes \partial_L \ = \ 0\;. 
\ee
As in double field theory, this constraint is meant to hold on any fields, parameters, 
and their products. This constraint has non-trivial 
solutions, which break  E$_{8(8)}$ to GL$(8)$ or GL$(7)\times {\rm SL}(2)$,
for which the EFT reduces to $D=11$ supergravity or type IIB, respectively,
for appropriate reformulations of these theories, 
as pioneered in~\cite{Nicolai:1986jk,Koepsell:2000xg} for E$_{8(8)}$.

The bosonic field content of the E$_{8(8)}$ EFT is given by
 \be\label{fieldcontent}
  \big\{\,e_{\mu}{}^{a}\,,\;{\cal V}_{M}{}^{\underline{M}}\,,\;A_{\mu}{}^{M}\,, \;B_{\mu\,M}\,\big\}\;.
 \ee 
It incorporates an external frame field (`dreibein') $e_{\mu}{}^{a}$, $\mu=0, 1, 2$, 
and an internal generalized frame field (`zweihundertachtundvierzigbein') ${\cal V}_{M}{}^{\underline{M}}$,
$M=1, \dots, 248$, 
parametrizing the coset space 
 E$_{8(8)}/$SO$(16)$. From the latter, we may construct the 
`generalized metric' as ${\cal M}_{MN}=({\cal V}{\cal V}^{T})_{MN}$. Crucially, the theory also requires the presence 
 of generalized gauge connections $A_{\mu}{}^{M}$ and $B_{\mu\,M}$,
  in order to consistently describe the complete degrees of freedom and dynamics
 of $D=11$ supergravity (necessarily including also some of the dual fields).  
 The theory is invariant under gauge symmetries with~parameters $\Lambda^M$, $\Sigma_M$, 
 acting as
\be\label{GenLieIntro}
\mathbb{L}_{(\Lambda,\Sigma)}V^M \ \equiv \ \Lambda^K\partial_K V^M
  -60\,\mathbb{P}^M{}_N{}^K{}_L \,\partial_K \Lambda^L \,V^N
  +\lambda\,\partial_N \Lambda^N\,V^M  - {\Sigma}_L f^{LM}{}_{N} V^N \;,
 \ee 
 on a vector $V^M$ of weight $\lambda$. The $\Lambda^M$ transformations generate the 
 generalized diffeomorphisms on the 248-dimensional space, following the definition for the
 smaller exceptional groups~\cite{Coimbra:2011ky} with $\mathbb{P}$ denoting the projector 
 onto the adjoint representation.
 The $\Sigma_M$ gauge symmetry is a new feature of the E$_{8(8)}$ EFT and describes
 a separate E$_{8(8)}$ gauge symmetry, however, with parameters $\Sigma_M$ that are 
 `covariantly constrained'. This means that they obey 
the same algebraic constraints as the derivatives in (\ref{secConstrIntro}), for instance
$\mathbb{P}_{MN}{}^{KL}\Sigma_K\otimes \partial_L=0$, etc..  
As a result, most of the components vanish after explicitly solving the section constraints, and
the E$_{8(8)}$ gauge symmetry is much smaller than is apparent from 
(\ref{GenLieIntro}) (as it should be, for otherwise all fields encoded in the E$_{8(8)}/$SO$(16)$ 
coset space would be pure gauge). 
This additional gauge symmetry is necessary for consistency. For instance, the generalized diffeomorphisms
in (\ref{GenLieIntro}) with parameter~$\Lambda^M$ do not close into themselves which has been
recognized as an obstacle in~\cite{Coimbra:2011ky,Berman:2012vc}.  They do however close in presence 
of the additional covariantly constrained gauge symmetry that constitutes a separate invariance of the theory. 
In other words, invariance of an action under generalized diffeomorphisms $\Lambda^M$ implies its invariance 
under further $\Sigma_M$ gauge transformations, as we shall explicitly confirm.
This type of gauge structure has first been revealed in the baby example of an ${\rm SL}(2)$ covariant formulation
of four-dimensional Einstein gravity~\cite{Hohm:2013jma}.

The constraints on the gauge parameter $\Sigma_M$ imply that also the associated connection 
$B_{\mu\,M}$ is covariantly constrained in the same sense, i.e.~it satisfies 
$\mathbb{P}_{MN}{}^{KL} B_{\mu\,K} \otimes \partial_L=0$, etc.. 
Such covariantly constrained compensating gauge fields are a generic feature of the exceptional
field theories and show up among the $(D-2)$-forms
(with $D$ counting the number of external dimensions). Therefore in $D=5$, these fields do not even
enter the Lagrangian~\cite{Hohm:2013vpa}, in $D=4$ they appear among the 2-forms
with St\"uckelberg coupling to the Yang-Mills field strengths~\cite{Hohm:2013uia}, 
while in $D=3$ they feature among the vector fields and thus directly affect the algebra 
of gauge transformations~(\ref{GenLieIntro}). In all cases, these constrained
gauge fields are  related to the appearance of the 
dual gravitational degrees of freedom as we discuss shortly.

The full E$_{8(8)}$ covariant action is given by
  \be
  S \ = \ \int d^3x \, d^{248}Y \,e\left( \widehat{R}+e^{-1}{\cal L}_{\rm CS}
  +\frac{1}{240}\,g^{\mu\nu}
  { D}_{\mu}{\cal M}^{MN}{ D}_{\nu}{\cal M}_{MN} -\,V({\cal M},g) \right)\;, 
  \label{fullaction0}
 \ee 
 and closely resembles the structure of three-dimensional gauged supergravities~\cite{Nicolai:2000sc}.
The various terms comprise a (covariantized) Einstein-Hilbert term, 
a Chern-Simons-type term for the gauge vectors, 
a covariantized kinetic term for the E$_{8(8)}/$SO$(16)$ coset fields, and a `potential'~$V$. 
The Chern-Simons term is a topological term that is needed to 
ensure the proper on-shell duality relations between `scalars' and `vectors'. 
The potential depends only on `internal' derivatives $\partial_M$ and can be written in 
a manifestly E$_{8(8)}$ covariant form as follows
 \bea\label{PotIntro}
  V({\cal M},g) & = &
  -\frac{1}{240}{\cal M}^{MN}\partial_M{\cal M}^{KL}\,\partial_N{\cal M}_{KL}+
  \frac{1}{2}{\cal M}^{MN}\partial_M{\cal M}^{KL}\partial_L{\cal M}_{NK} \\
  &&{}
  +\frac1{7200}\,f^{NQ}{}_P f^{MS}{}_R\,
  {\cal M}^{PK} \partial_M {\cal M}_{QK} {\cal M}^{RL} \partial_N {\cal M}_{SL} 
  \nonumber\\
  &&{}
  -\frac{1}{2}g^{-1}\partial_Mg\,\partial_N{\cal M}^{MN}-\frac{1}{4}  {\cal M}^{MN}g^{-1}\partial_Mg\,g^{-1}\partial_Ng
  -\frac{1}{4}{\cal M}^{MN}\partial_Mg^{\mu\nu}\partial_N g_{\mu\nu}\;. 
  \nonumber
\eea
Its form is determined such that it leads to a gauge invariant action both w.r.t.~the $\Lambda^M$ 
and $\Sigma_M$ gauge transformations of (\ref{GenLieIntro}). 
Previous attempts to construct an E$_{8(8)}$ covariant formulation (of truncations of $D=11$ supergravity) 
missed the second line of (\ref{PotIntro}) involving the explicit  E$_{8(8)}$ structure constants $f^{MN}{}_K$ 
\cite{Godazgar:2013rja}. This term is indispensable for gauge invariance of the potential $V$
and for the match with $D=11$ supergravity as we shall explain. 
All four terms in the action (\ref{fullaction0}) are seperately 
gauge invariant w.r.t.~$\Lambda$ and $\Sigma$, but the theory is also invariant under 
non-manifest external diffeomorphisms of the $x^{\mu}$, generated by  a parameter $\xi^{\mu}(x,Y)$.
This symmetry fixes all the relative coefficients in (\ref{fullaction0}),
such that this is the unique two-derivative action with all the required symmetries.

We close the introduction by a discussion of how the above EFT resolves what is often 
referred to as the `dual graviton problem'. 
This problem comes about because the  E$_{8(8)}$ coset representative ${\cal M}_{MN}$ depends on 
components $\varphi_m$, $m=1,\ldots, 8$, that in three dimensions
are dual to the Kaluza-Klein vectors $A_{\mu}{}^{m}$. 
As the latter originate from components of the $D=11$ metric, this amounts to including in the theory
components of a `dual graviton'~\cite{Curtright:1980yk,Hull:2000zn,West:2001as,Hull:2001iu} 
at the full non-linear level, something that is considered impossible on the grounds of the no-go theorems
in \cite{Bekaert:2002uh,Bekaert:2004dz}.  In EFT this problem is resolved due to the presence of the extra 
E$_{8(8)}$ gauge symmetry from (\ref{GenLieIntro}). 
Solving the section constraints (\ref{secConstrIntro}) such that the 
theory reduces to $D=11$ supergravity, this covariantly constrained gauge symmetry 
reduces to a St\"uckelberg shift symmetry with $8$ parameters, which can be used 
to gauge away all the dual graviton components $\varphi_m$. Consequently, in the gauge invariant 
potential (\ref{PotIntro}) all components  $\varphi_m$ drop out upon solving the 
section constraint, which is necessary for the theory to match $D=11$ supergravity. 
The same conclusions hold for the solution corresponding to 
type IIB. 
Let us finally note that although the dual graviton components $\varphi_m$ are pure gauge 
for the $D=11$ and $D=10$ solutions, once we consider strict dimensional reduction to $D=3$, 
the $\varphi_m$ are propagating fields among the scalars of the E$_{8(8)}/$SO$(16)$ coset space. 
Indeed, in this case the Chern-Simons term implies that $B_{\mu\,m}$ is pure gauge, so the 
extra gauge symmetry can be fixed by gauging $B_{\mu\,m}$ away.

This paper is organized as follows. In sec.~\ref{sec2} we introduce the E$_{8(8)}$ generalized 
Lie derivatives and the covariantly constrained E$_{8(8)}$ gauge symmetry. Next, we introduce gauge 
vectors for these symmetries and define covariant derivatives and field strengths. 
In sec.~\ref{sec3} we present the various terms in the action and prove its gauge invariance
under internal and external diffeomorphisms. 
In particular, we fix all relative coefficients in the action (\ref{fullaction0})
by requiring invariance under external diffeomorphisms. 
Finally, in sec.~\ref{sec4}, we discuss the match with $D=11$ supergravity and type IIB. 
Specifically, we discuss how the dual graviton (problem) disappears. 
We conclude in sec.~\ref{sec5}. Some details on the proof of closure of the E$_{8(8)}$ generalized 
Lie derivatives are presented in the appendix.

\section{E$_{8(8)}$ Gauge Structure}\label{sec2}
In this section we introduce E$_{8(8)}$ covariant generalized Lie derivatives, 
which close according to an E-bracket, up to a separate `covariantly constrained' E$_{8(8)}$ 
gauge symmetry. This mean that the E$_{8(8)}$ gauge parameter is subject to the same 
section constraints as the extended derivatives. Then we introduce gauge fields $A_{\mu}{}^{M}$ for 
the E-bracket and covariantly constrained gauge fields $B_{\mu M}$ for E$_{8(8)}$.

\subsection{E$_{8(8)}$ generalized Lie derivatives}
We start by recalling a few generalities of E$_{8(8)}$. Its Lie algebra is 248-dimensional,
and the adjoint representation is the smallest fundamental representation. 
We denote the generators by $(t^M)^N{}_{K}=-f^{MN}{}_{K}$, 
with structure constants $f^{MN}{}_{K}$ and adjoint indices $M,N=1,\ldots, 248$. 
The maximal compact subgroup is SO$(16)$, under which E$_{8(8)}$ decomposes 
as ${\bf 248}\rightarrow {\bf 120}\oplus {\bf 128}$. There is an invariant symmetric tensor $\eta_{MN}$, 
the Cartan-Killing form, 
which we normalize by 
 \be\label{CartanKilling}
  \eta^{MN} \ = \  \frac{1}{60}\,{\rm tr}\big(t^M t^N\big) \ = \ \frac{1}{60} f^{MK}{}_{L} f^{NL}{}_{K}\;, 
 \ee 
and which we freely use to raise and lower adjoint indices.  
Given this invariant metric, the tensor product of the adjoint with the co-adjoint representation is equivalent
 to  ${\bf 248}\otimes {\bf 248}$
and decomposes as follows 
 \be\label{248squared}
  {\bf 248}\otimes {\bf 248} \ \rightarrow \ {\bf 1}\oplus  {\bf 248}\oplus {\bf 3875}\oplus {\bf 27000}
  \oplus {\bf 30380}\;.
 \ee 
In particular, it contains the adjoint representation, and in the following 
we need the corresponding projector: 
\begin{eqnarray}
\mathbb{P}^M{}_N{}^K{}_L &=&
\frac1{60}\,f^M{}_{NP}\,f^{PK}{}_L{} \\ 
&=&
\frac1{30}\, \delta_{(N}^M \delta_{L)}^K 
-\frac7{30}  (\mathbb{P}_{3875}){}^{MK}{}_{NL}
-\frac1{240}\,\eta^{MK}\eta_{NL}
+\frac1{120}\, f^{MK}{}_{P}\,f^{P}{}_{NL}{}
\;.\nonumber 
\label{pex}
\end{eqnarray}
Here we used eqs.~(2.15) in \cite{Koepsell:1999uj}, and the projector onto the ${\bf 3875}$ which is given by
  \be\label{3875proj}
   (\mathbb{P}_{3875}){}^{MK}{}_{NL} \ = \ \frac{1}{7}\,\delta^M_{(N}\, \delta^{K}_{L)}
   -\frac{1}{56}\,\eta^{MK}\,\eta_{NL}-\frac{1}{14}\,f^{P}{}_{N}{}^{(M}\, f_{PL}{}^{K)}\;. 
  \ee 
We refer to \cite{Koepsell:1999uj,Koepsell:2000xg} for other useful E$_{8(8)}$ identities. 

Let us now discuss the generalized spacetime and geometry based on E$_{8(8)}$. We introduce $248$ 
coordinates $Y^M$ in the adjoint representation, but we subject all functions (i.e.~including all fields and 
gauge parameters and all their products) to the covariant section constraints~(\ref{secConstrIntro}). 
These are necessary in order for the symmetries of the theory to close into an algebra. These symmetries
comprise generalized diffeomorphisms on the 248-dimensional space, 
together with a covariantly constrained E$_{8(8)}$ gauge symmetry. 
Specifically, denoting by $\Lambda^M$ and $\Sigma_M$ the parameters for 
generalized diffeomorphisms and constrained E$_{8(8)}$, respectively, 
we define the generalized Lie derivative on a vector by 
 \bea
  \delta V^M &=& \mathbb{L}_{(\Lambda,\Sigma)}V^M \ \equiv \ \Lambda^K\partial_K V^M
  -60\,\mathbb{P}^M{}_N{}^K{}_L \,\partial_K \Lambda^L \,V^N
  +\lambda(V)\,\partial_N \Lambda^N\,V^M \nonumber\\
  &&{} \qquad\qquad\qquad -{\Sigma}_L f^{LM}{}_{N} V^N 
  \;. 
 \label{DV}
 \eea
Analogously, one may define the generalized Lie derivative acting  on tensors 
with an arbitrary number of adjoint E$_{8(8)}$ indices.  
The first line of (\ref{DV}) defines the generalized Lie derivative w.r.t.~$\Lambda^M$, 
in accordance with the definition for the smaller exceptional groups~\cite{Coimbra:2011ky,Berman:2012vc},
where we also allowed for a general density weight $\lambda$. The second line is a novel feature of the 
E$_{8(8)}$ EFT. It defines the 
covariantly constrained E$_{8(8)}$ action, i.e.\ describes an E$_{8(8)}$ rotation with a parameter $\Sigma_M$
which itself satisfies the same algebraic conditions (\ref{secConstrIntro}) as the partial derivatives.
Concretely,  we require that
 \be
  \big(\mathbb{P}_{1+248+3875}\big)_{MN}{}^{KL} C_K\otimes C'_L \ = \ 0\;,
  \qquad \text{for} \quad C_M\,, C_M' \ \in \ \{\partial_M, {B}_{\mu\, M},{\Sigma_M}\}\;,
  \label{section_full}
 \ee 
where ${B}_{\mu\, M}$ denotes the gauge connection associated to the $\Sigma_M$ symmetry of (\ref{DV}). 
This means that for any expression containing two objects, $C_M$ and $C_N'$, from the list above,  
the part in the tensor product that is projected out by this constraint can be consistently set to zero. 
Explicitly, we have for the individual irreducible representations, 
 \be\label{secconstr}
  \eta^{MN}C_M\otimes  C'_N \ = \ 0 \;, \quad 
  f^{MNK}C_N\otimes C'_K \ = \ 0\;, \quad 
  \big(\mathbb{P}_{3875}\big)_{MN}{}^{KL} C_K\otimes C'_L \ = \ 0
  \;.
 \ee 
This implies in particular $\eta^{MN}\partial_M\partial_NA=\partial^M\partial_MA=0$, but also 
$\partial^MA\,\partial_M B=0$, for arbitrary functions $A,B$, and relations like $f^{MNK}B_{\mu N}\, \partial_KA=0$ 
involving the covariantly constrained gauge field $B_{\mu\, M}$. These relations imply that for any solution of the 
section constraint only a subset of coordinates among the $Y^M$ survives, while also only the 
`corresponding' components of $B_{\mu\, M}$ are present, as we will explain in more detail below.    

Before determining the gauge algebra satisfied by (\ref{DV}) we briefly discuss that the above 
gauge transformations (\ref{DV}) possess `trivial' gauge parameters. For these parameters 
the action of the associated generalized Lie derivative on any field vanishes by virtue of the 
section constraints (\ref{secconstr}). 
The following parameters are trivial in this sense, 
 \bea
 \Lambda^M &\equiv&\eta^{MN}\Omega_{N}\;,\qquad \text{$\Omega_N\;$ covariantly constrained \`a la (\ref{section_full})}\;, 
 \nonumber\\
 \Lambda^M &\equiv& (\mathbb{P}_{3875}){}^{MK}{}_{NL}\,\partial_K \chi^{NL}
 \;.
 \label{triv1}
 \eea
Here, in the first line, $\Omega_N$ is covariantly constrained in the sense that it 
satisfies the same constraints as the $C_N$ in (\ref{section_full}), (\ref{secconstr}). 
E.g.\ choosing $\Omega_N=\partial_N\chi$
we infer that $\Lambda^M=\partial^M\chi$ is a trivial parameter, in analogy to DFT. 
For the first parameter in (\ref{triv1}) it is straightforward to see with (\ref{pex}) and the constraints 
(\ref{secconstr}) that the generalized Lie derivative (\ref{DV}) is zero on fields. 
As an illustration for the use of constraints, we prove explicitly the triviality of the 
second parameter in (\ref{triv1}). We first note that in this case the transport term and density term 
(i.e.~the first and third term) in (\ref{DV}) immediately vanish as a consequence of the third 
constraint in (\ref{secconstr}). Thus, the action of the generalized Lie derivative reads 
 \be\label{TRIVcomp}
 \begin{split}
  \mathbb{L}_{\Lambda} V^M 
  \ &= \ -60\, \mathbb{P}^{M}{}_{N}{}^{(P}{}_{Q}(\mathbb{P}_{3875})^{R)Q}{}_{ST}\,\partial_P\partial_R\chi^{ST} V^N\\
  \ &= \ - f^{M}{}_{NX}\big(f^{X(P}{}_{Q}(\mathbb{P}_{3875})^{R)Q}{}_{ST}\big)\,\partial_P\partial_R\chi^{ST} V^N\;. 
 \end{split}
 \ee 
Next, we use that $\mathbb{P}_{3875}$ is an invariant tensor under the adjoint 
action of E$_{8(8)}$, as is manifest from its definition (\ref{3875proj}). This means 
 \be
  f^{X(P}{}_{Q}(\mathbb{P}_{3875})^{R)Q}{}_{ST} - f^{XQ}{}_{(S}(\mathbb{P}_{3875})^{PR}{}_{T)Q} \ = \ 0\;.
 \ee 
Thus, we can replace the structure in (\ref{TRIVcomp}) by the second term in here. Being 
contracted with $\partial_P\partial_R$ it then follows from the third constraint in (\ref{secconstr}) 
that this vanishes, completing the proof that the associated generalized Lie derivative acts trivially. 
 
Next, we discuss a novel phenomenon for the E$_{8(8)}$ case: there are combinations of parameters 
$\Lambda$ and $\Sigma$ whose combined action is trivial on all the fields.
Specifically, the generalized Lie derivative (\ref{DV}) with parameters
 \bea\nonumber 
 \Lambda^M &=& f^{MN}{}_{K}\Omega_{N}{}^{K}\;, \qquad \text{$\Omega_{N}{}^{K}$\;
 covariantly constrained in first index}\;, \\ 
 \Sigma_M  &=&   \partial_M\Omega_{N}{}^{N}+\partial_N\Omega_M{}^{N}\;, 
 \label{triv2}
 \eea
acts trivially for a general tensor $\Omega_{N}{}^{K}$ that is covariantly constrained in the first index
in the sense of (\ref{section_full}), (\ref{secconstr}). 
An example is given by $\Omega_{M}{}^{N}=\partial_M\chi^N$ with arbitrary $\chi^N$, 
so we conclude as a special case of (\ref{triv2}) that 
\be\label{specialTRIV}
 \Lambda^M \, \equiv \, f^{MN}{}_{K} \partial_N \chi^K \;,
 \qquad \Sigma_M ~\equiv~ 2\partial_M\partial_N\chi^N\;, 
\ee
has trivial action on all the fields.  
In order to verify the triviality of (\ref{triv2}) let us first prove the following useful 
Lemma:
 \be\label{Lemma}
  f^{P}{}_{M}{}^{K}\,f_{PN}{}^{L}\,C_{K}\otimes C'_{L} \ = \ C_M\otimes C'_N+C_N\otimes C'_M\;, 
 \ee
for any covariantly constrained objects $C_M$, $C'_M$.  To prove this we compute
 \be
  \begin{split}
   f^{P}{}_{M}{}^{K}\,f_{PN}{}^{L}\,C_{K}\otimes C'_{L} \ &= \ 
    \big(f^{P}{}_{M}{}^{[K}\,f_{PN}{}^{L]}+f^{P}{}_{M}{}^{(K}\,f_{PN}{}^{L)}\big)\,C_{K}\otimes C'_{L} \\
    \ &= \ \big(-\tfrac{1}{2}f^{PKL} f_{PNM}+2\delta_{(M}^K\,\delta_{N)}^L\\
    &\qquad \, -\tfrac{1}{4}\eta_{MN}\eta^{KL}
    -14 (\mathbb{P}_{3875})^{KL}{}_{MN}\big)\,C_{K}\otimes C'_{L}\\
    \ &= \ C_M\otimes C'_N+C_N\otimes C'_M\;. 
 \end{split}   
 \ee    
In the second line we used  the Jacobi identity and rewrote the symmetrized $ff$ term in 
terms of the ${\bf 3875}$ projector (\ref{3875proj}). In the final step we used the section constraints 
(\ref{secconstr}). This completes the proof of (\ref{Lemma}). 
It is now straightforward to verify the triviality of (\ref{triv2}). First, the transport and density terms 
vanish immediately as a consequence of the second constraint in (\ref{secconstr}). 
The remaining projector term, in the first form of the projector in (\ref{pex}), can then be 
simplified by (\ref{Lemma}) to show that this cancels the $\Sigma$ terms from (\ref{triv2}). 
Another immediate consequence of (\ref{Lemma}) is that for a generalized vector $\Omega_M$ 
(of weight zero) that is covariantly constrained, the generalized Lie derivative reduces to 
\bea
\delta_{\Lambda} \Omega_M &=& 
 \Lambda^N \partial_N \Omega_M +   \partial_N \Lambda^N \Omega_M
+  \partial_M \Lambda^N \Omega_N
\;, 
\label{act_con}
\eea
which will be used below. 

We close this section by discussing closure of the gauge transformations. In contrast to the analogous structures 
for E$_{n(n)}$ with $n\leq 7$, the generalized Lie derivatives do not close by themselves, but only up to 
(constrained) local E$_{8(8)}$ gauge transformations. Specifically, one finds closure   
 \bea\label{closure}
  \big[\,\delta_{(\Lambda_{1},\Sigma_1)},\,\delta_{(\Lambda_{2},\Sigma_2)}\,\big] &=& 
  \delta_{[(\Lambda_{2},\Sigma_2),(\Lambda_{1},\Sigma_1)]_{\rm E}}\;, 
 \qquad
 [(\Lambda_{2},\Sigma_2),(\Lambda_{1},\Sigma_1)]_{\rm E} ~\equiv~ (\Lambda_{12},\Sigma_{12})
 \;,\quad
 \eea
with the effective parameters
 \bea\label{EFFimp}
  \Lambda^M_{12} &\equiv&
2\,\Lambda_{[2}^N\partial_N \Lambda_{1]}^M
-14\,  (\mathbb{P}_{3875}){}^{MK}{}_{NL} \,\Lambda_{[2}^N \partial_K \Lambda_{1]}^L
-\frac1{4}\,\eta^{MK}\eta_{NL}\,\Lambda_{[2}^N \partial_K \Lambda_{1]}^L\nonumber\\
&&{}
+\frac1{4}\,f^{MN}{}_{P}\,\partial_N(f^{P}{}_{KL}\Lambda_2^K  \Lambda_1^L)
\;,\nonumber\\[1ex]
\Sigma_{12\,M} &\equiv&
-2\,\Sigma_{[2\,M} \partial_N \Lambda_{1]}^N 
+2\,\Lambda_{[2}^N \partial_N \Sigma_{1]\,M} 
-2\, \Sigma_{[2}^N \partial_M \Lambda_{1]\,N}
+f^N{}_{KL} \,\Lambda_{[2}^K\,\partial_M\partial_{N}\Lambda_{1]}^L
\;.\qquad
 \eea 
Note that here is an ambiguity in the form of the effective gauge parameters, because they can be 
redefined by trivial gauge parameters,  (\ref{triv1}) or (\ref{triv2}), without spoiling closure. 
In particular, the term in the second line of $\Lambda_{12}$ could have been dropped, 
using (\ref{triv2}), at the cost of extra terms in $\Sigma_{12}$. The form here has been chosen for later convenience.  
We stress again that closure only holds because of the separate (covariantly constrained) E$_{8(8)}$ 
gauge symmetry. Note that this is a rather non-trivial statement, because the effective $\Sigma_{12}$
parameter needs to be compatible with the covariant section constraints (\ref{secconstr}). The compatibility is manifest from 
the form in (\ref{EFFimp}), because  in each term the free index $M$ is carried by a constrained object, 
$\Sigma_M$ or $\partial_M$. As this interplay between generalized diffeomorphisms and a 
separate but constrained gauge symmetry is somewhat unconventional we prove gauge closure (\ref{closure}), 
(\ref{EFFimp}) explicitly in the appendix. 
We finally note that the gauge algebra of $\Sigma$ transformations with themselves 
is abelian, for the effective parameter $\Sigma_{12}^M=f^{MNK}\Sigma_{2N}\Sigma_{1K}$ is actually 
zero by the section constraints~(\ref{secconstr}).

\subsection{Gauge fields for E$_{8(8)}$ E-bracket}
We now introduce gauge fields for the local symmetries generated by $\Lambda^M$ and $\Sigma_M$. 
Specifically, these parameters are functions of $x^{\mu}$ and $Y^M$, requiring in particular covariant 
derivatives $D_{\mu}$ for the external coordinates. Denoting the gauge fields for the $\Lambda^M$
symmetries by $A_{\mu}{}^{M}$ and those for the $\Sigma_M$ symmetries by $B_{\mu M}$, 
the covariant derivative on any tensor with an arbitrary number of adjoint E$_{8(8)}$ indices is defined by 
 \be\label{covder}
  D_{\mu} \ \equiv \ \partial_{\mu}-\mathbb{L}_{(A_{\mu},B_{\mu})}\;, 
 \ee 
where the generalized Lie derivative ${\mathbb L}$ acts according to the representation the tensor field lives in. 
For instance, using (\ref{DV}) one finds its action on a vector of zero weight 
 \be
  D_{\mu}V^M \ = \ \partial_{\mu}V^M- A_{\mu}{}^K\partial_K V^M
  +60\,\mathbb{P}^M{}_N{}^K{}_L \,\partial_K A_{\mu}{}^L \,V^N
  +B_{\mu}{}^L f^{M}{}_{NL} V^N
 \;.
 \ee
The transformation rules for $A$ and $B$ are determined by the requirement that 
the covariant derivatives (\ref{covder}) transform covariantly. 
In general, their gauge transformations can be computed from  
 \be
 \begin{split}
  (\delta_{(\Lambda,\Sigma)} A,\delta_{(\Lambda,\Sigma)}B) \ &\equiv \ 
   (\partial\Lambda ,\partial \Sigma) + 
   \big[(\Lambda,\Sigma),(A,B)\big]_{\rm E}\;, 
 \end{split}
 \ee
with the E-bracket defined by (\ref{closure}).  
Using (\ref{EFFimp}) one computes for the components 
 \be\label{STEPGauge}
 \begin{split}
  \delta_{(\Lambda,\Sigma)}  A_{\mu}{}^M \ = \  & D_{\mu}\Lambda^M -\partial_N A_{\mu}{}^N\,\Lambda^M
 +7\,\mathbb{P}_{3875}{}^{MN}{}_{KL}\big(\Lambda^K\partial_NA_{\mu}{}^L+A_{\mu}{}^K\partial_N\Lambda^L\big)\\
 &- B_{\mu}{}^L f^M{}_{NL}\,\Lambda^N
 +\frac{1}{8}\eta^{MN}\eta_{KL}\big(\Lambda^K\partial_NA_{\mu}{}^L+A^K\partial_N\Lambda^L\big)\\
  &+\frac{1}{4}f^{MN}{}_{P}f^{P}{}_{KL} \, \big(\partial_N\Lambda^K A_{\mu}{}^L-\Lambda^K \partial_N A_{\mu}{}^L \big)\;, \\[1ex]
  \delta_{(\Lambda,\Sigma)}  B_{\mu M} \ = \ \; &
  D_{\mu}\Sigma_M + \partial_N\big(B_{\mu M}\Lambda^N\big) +B_{\mu}{}^N\partial_M\Lambda_N\\
&+\frac12\,f^{N}{}_{KL}\big(  \Lambda^K \partial_M\partial_N  A_{\mu}{}^L-
  A_{\mu}{}^K \partial_M\partial_N  \Lambda^L \big) \;.
  \end{split}
 \ee 
Note in particular, that the gauge field $A$ and its parameter $\Lambda$ carries weight one
(and we have explicitly spelled out the weight term in (\ref{STEPGauge})),
whereas $B$ and $\Sigma$ carry weight 0. 

Because of the existence of trivial gauge parameters,  c.f.~(\ref{triv1}) and (\ref{triv2}) 
discussed in the previous subsection, 
the gauge transformations of $A$ and $B$ are determined from the covariance of (\ref{covder}) 
only up to redefinitions by trivial parameters.
Specifically, the covariant derivatives (\ref{covder}) are invariant under the
 following vector shift transformations
 \bea
 \delta A_\mu{}^M &=& 
  \partial^M \Xi_\mu
+ (\mathbb{P}_{3875}){}^{MK}{}_{NL}\,\partial_K \Xi_\mu{}_{3875}^{NL}
+ f^{MN}{}_{K}\Xi_{\mu}{}_{N}{}^{K}\;, \nonumber\\
 \delta B_\mu{}_M &=& \partial_M\Xi_{\mu}{}_{N}{}^{N}+\partial_N\Xi_{\mu}{}_M{}^{N}
  \;,
  \label{shiftAB}
 \eea
 with $\Xi_{\mu}{}_M{}^{N}$ constrained in the first index.
We now redefine the gauge transformations of $A$ and $B$ by adding trivial gauge transformations 
of this form, with parameters  
 \be
  \begin{split}
   \Xi_{\mu}{}^{KL}_{3875} \ &= \ -7(\mathbb{P}_{3875})^{KL}{}_{PQ} A_{\mu}{}^{P}\Lambda^Q\;,  \\
   \Xi_{\mu} \ &= \ -\frac{1}{8} A_{\mu}{}^K\Lambda_K\;, \\
   \Xi_{\mu N}{}^{K} \ &= \ -B_{\mu N}\Lambda^K+\frac{1}{4}f^{K}{}_{PQ}\Lambda^{P}\partial_NA_{\mu}{}^Q
   -\frac{1}{4}f^{K}{}_{PQ}\partial_N\Lambda^P A_{\mu}{}^Q\;. 
  \end{split}
 \ee  
The gauge transformations (\ref{STEPGauge}) then take the more compact form
 \be\label{gaugeAB}
  \begin{split}
   \delta A_{\mu}{}^M \ &= \ D^{(1)}_{\mu}\Lambda^M\;, \\
   \delta B_{\mu M} \ &= \ D^{(0)}_{\mu}\Sigma_M-\Lambda^N\partial_M B_{\mu N}+f^{N}{}_{KL} \Lambda^K\partial_M\partial_NA_{\mu}{}^L\;,
  \end{split}
 \ee 
where we have indicated the respective weights by the superscripts $D_\mu^{(\lambda)}$\,.
This is the final form of the gauge transformations that we use in the following. 

Let us now turn to the definition of gauge covariant curvatures or field strengths. 
Part of these curvatures can be read off from the commutator of 
covariant derivatives, 
\bea
[{D}_\mu,D_\nu] \, V^M &=&
-\mathbb{L}_{(F_{\mu\nu},G_{\mu\nu})} V^M
\;.
\eea
More precisely, this determines the field strengths up to trivial terms that 
drop out of the generalized Lie derivatives, for which we find 
\bea
   F_{\mu\nu}{}^{M} 
   &= &2\,\partial_{[\mu}A_{\nu]}{}^{M}-2A_{[\mu}{}^{N}\partial_NA_{\nu]}{}^{M}+14\,(\mathbb{P}_{3875}){}^{MN}{}_{KL}
   A_{[\mu}{}^{K}\partial_N A_{\nu]}{}^{L}+\frac{1}{4}A_{[\mu}{}^{N}\partial^M A_{\nu]N}\nonumber\\
   &&{}-\frac{1}{2}f^{MN}{}_{P}f^{P}{}_{KL}A_{[\mu}{}^{K}\partial_N A_{\nu]}{}^{L}
  \;,\nonumber\\
  G_{\mu\nu M} &=& 2\,D_{[\mu}B_{\nu]M}-f^{N}{}_{KL}A_{[\mu}{}^{K}\partial_M\partial_NA_{\nu]}{}^{L}
\;.
\eea  
These field strengths do not transform covariantly, but the failure of covariance is 
of a `trivial' form that can be compensated by adding two-form couplings and assigning to them 
appropriate gauge transformations in the general spirit of the $p$-form tensor hierarchy~\cite{deWit:2008ta}. 
We thus introduce the fully covariant curvatures 
\bea
  {\cal F}_{\mu\nu}{}^M &\equiv& 
F_{\mu\nu}{}^{M}+14\,(\mathbb{P}_{3875}){}^{MN}{}_{KL}\,\partial_N
  C_{\mu\nu}{}_{(3875)}^{KL}+\frac{1}{4}\partial^M C_{\mu\nu}
  +2f^{MN}{}_{K} C_{\mu\nu N}{}^{K}\;,
  \nonumber\\
 {\cal G}_{\mu\nu M} &=&  G_{\mu\nu M}+2\, \partial_N  C_{\mu\nu}{}_M{}^{N}+2\, \partial_M C_{\mu\nu N}{}^{N}  
\;, 
\eea
with two-form fields $C_{\mu\nu}{}_{(3875)}^{KL}$, $C_{\mu\nu}$, and $C_{\mu\nu\, M}{}^{N}$,
where as in (\ref{triv2}) the two-form $C_{\mu\nu\, M}{}^{N}$ is covariantly constrained in the first index. 
The general variation of these curvatures takes a covariant form, 
\bea\label{FGC}
   \delta {\cal F}_{\mu\nu}{}^{M} &=& 2D^{(1)}_{[\mu}\delta A_{\nu]}{}^{M}+14\,(\mathbb{P}_{3875}){}^{MN}{}_{KL}\,\partial_N
  \Delta C_{\mu\nu}{}_{(3875)}^{KL}+\frac{1}{4}\partial^M\Delta C_{\mu\nu}
  +2f^{MN}{}_{K}\Delta C_{\mu\nu N}{}^{K}\;,
  \nonumber\\
 \delta {\cal G}_{\mu\nu}{}_{M} &=&2D^{(0)}_{[\mu}\,\delta B_{\nu]}{}_{M}- 2\partial_MB_{[\mu}{}^{N}\,\delta A_{\nu] N}
-2\,f^{N}{}_{KL} \delta A_{[\mu}{}^{K}\partial_M\partial_N A_{\nu]}{}^{L}
 +2\, \partial_N\Delta C_{\mu\nu}{}_M{}^{N}
   \nonumber\\
   &&{} 
   +2\,\partial_M\Delta C_{\mu\nu N}{}^{N}
   \;,
\eea  
where we defined the covariant variations 
\bea
\Delta C_{\mu\nu}{}_{(3875)}^{KL} &\equiv& \delta C_{\mu\nu}{}_{(3875)}^{KL}+A_{[\mu}{}^{K}\delta A_{\nu]}{}^{L}
\;,\nonumber\\
\Delta C_{\mu\nu}&\equiv&\delta C_{\mu\nu}+A_{[\mu}{}^{K}\delta A_{\nu]}{}_{K}
\;,\nonumber\\
\Delta C_{\mu\nu N}{}^{K}&\equiv&\delta C_{\mu\nu N}{}^{K}+
B_{[\mu N}\delta A_{\nu]}{}^{K}
  -\frac{1}{4}f^{K}{}_{PQ}\big(A_{[\mu}{}^{P}\partial_N\delta A_{\nu]}{}^{Q}-\partial_NA_{[\mu}{}^{P}\delta A_{\nu]}{}^{Q}\big)
  \;.\quad
\eea 

We stress that although we had to introduce the additional two-forms in order to define gauge covariant curvatures,
all of them will eventually drop out from the action and the transformation rules. 
They can be viewed as a convenient tool that allows us to define the Lagrangian in a rather compact form in terms of 
manifestly covariant quantities whereas we could also have defined the Lagrangian directly in terms of the original fields
and confirmed its gauge invariance by an explicit computation.
The two-forms $C_{\mu\nu}$ and $C_{\mu\nu}{}_{(3875)}^{KL}$ already show up in the dimensionally reduced theory
upon extending on-shell the supersymmetry algebra and first order duality equations beyond the fields present in the 
Lagrangian~\cite{deWit:2008ta}.

We now specialize to the transformation of the curvatures under $\Lambda$ and $\Sigma$ 
gauge transformations (\ref{STEPGauge}). 
The field strength ${\cal F}_{\mu\nu}{}^M$ transforms covariantly in that 
\bea
   \delta_{\Lambda,\Sigma}{\cal F}_{\mu\nu}{}^{M} &=& \ \mathbb{L}_{(\Lambda,\Sigma)}{\cal F}_{\mu\nu}{}^{M}\;, 
\eea
with weight $\lambda=1$, provided the two-forms $C_{\mu\nu}$ transform as
\bea   
   \Delta C_{\mu\nu(3875)}{}^{KL} &=& {\cal F}_{\mu\nu}{}^{(K}\Lambda^{L)}\;, \nonumber\\ 
   \Delta  C_{\mu\nu} &=& {\cal F}_{\mu\nu}{}^{M}\Lambda_M\;, \nonumber\\ 
   \Delta C_{\mu\nu N}{}^{K} &=&  
   \frac{1}{4}f^{K}{}_{PQ}\big(\partial_N{\cal F}_{\mu\nu}{}^{P} \Lambda^Q-\partial_N\Lambda^Q{\cal F}_{\mu\nu}{}^{P}\big)
   +\frac{1}{2}{\cal G}_{\mu\nu N}\Lambda^K+\frac{1}{2}\Sigma_N{\cal F}_{\mu\nu}{}^{K}\;.
\eea 
On the other hand, the field strength ${\cal G}_{\mu\nu}{}^M$ transforms as
\bea
\delta_{\Lambda,\Sigma}{\cal G}_{\mu\nu M} &=& \ \mathbb{L}_{(\Lambda,\Sigma)}{\cal G}_{\mu\nu M}
-f^{N}{}_{KL}  {\cal F}_{\mu\nu}{}^{K}\, \partial_M\partial_N\Lambda^L\, 
  +\partial_M\Sigma_N\,{\cal F}_{\mu\nu}{}^{N}\;,
  \label{delG}
\eea 
where the generalized Lie derivative acts on a tensor of weight 0.
 These turn out to be the proper transformation rules in order to define
 a gauge invariant Chern-Simons term below.  To this end we will
 furthermore derive a set of generalized Bianchi identities (\ref{genBianchi})
satisfied by the curvatures ${\cal F}_{\mu\nu}{}^{M}$ and ${\cal G}_{\mu\nu M}$\,.


\section{The Action}\label{sec3}


With the structures set up in the previous section we are now in position
to define the various terms in the action (\ref{fullaction0})  
\be
  S \ = \ \int d^3x \, d^{248}Y \left({\cal L}_{\rm EH}+{\cal L}_{\rm CS}
  +{\cal L}_{\rm kin} -e\,V({\cal M},g) \right)\;.
  \label{fullaction}
 \ee 
We describe them one by one. 
We then verify that the action is invariant under generalized internal and 
properly defined external diffeomorphisms, which in turn fixes all the relative coupling constants.

 \paragraph{Einstein-Hilbert and kinetic term}

As in~\cite{Hohm:2013nja,Hohm:2013pua}, the Einstein-Hilbert term in (\ref{fullaction}) reads 
\bea\label{COvEH}
{\cal L}_{\rm EH} \ = \ e\widehat{R} \ \equiv \ eg^{\mu\nu}\widehat{R}_{\mu\nu}\;,
\eea
and is constructed from contraction of the 
improved Riemann tensor 
  \be
  \widehat{R}_{\mu\nu}{}^{ab} \ \equiv \  R_{\mu\nu}{}^{ab}[\omega]+{\cal F}_{\mu\nu}{}^{M}
  e^{\rho[a}\partial_M e_{\rho}{}^{b]}\,,
  \label{improvedR}
 \ee
where $R_{\mu\nu}{}^{ab}[\omega]$ denotes the covariantized curvature of the spin connection 
$\omega_\mu{}^{ab}$, 
which in turn is defined by the covariantized vanishing torsion condition
\bea
0 &=& {D}_{[\mu} e_{\nu]}{}^{a} ~\equiv~
\partial_{[\mu} e_{\nu]}{}^{a} -A_{[\mu}{}^K \partial_K  e_{\nu]}{}^{a} -\partial_K A_{[\mu}{}^K\,e_{\nu]}{}^{a}
+ \omega_{[\mu}{}^{ab}\,e_{\nu] b}
\;.
\label{omega}
\eea
In particular, the dreibein $e_\mu{}^a$ is an ${\rm E}_{8(8)}$ scalar-density of weight $\lambda=1$\,.
Note from the second form in (\ref{COvEH}) 
that with this weight the Einstein-Hilbert term has a total weight of $1$, as needed for local 
$\Lambda^M$ gauge invariance. 
The second term in (\ref{improvedR}) ensures covariance of the Riemann tensor under local Lorentz transformations.
As a result, the Einstein-Hilbert term ${\cal L}_{\rm EH}$ is invariant under local Lorentz transformations and internal generalized diffeomorphisms.
We note that the term is also invariant under the vector shift symmetries (\ref{shiftAB}), notably all two-form contributions 
in ${\cal F}_{\mu\nu}{}^M$ drop out from (\ref{improvedR}).

The matter sector of the theory comprises 128 `scalar' fields which 
as in the three-dimensional maximal theory~\cite{Marcus:1983hb} parametrize the coset space 
${\rm E}_{8(8)}/{\rm SO}(16)$.
In terms of the symmetric group-valued $248\times248$ matrix ${\cal M}_{MN}$
(and its inverse ${\cal M}^{MN}$), the kinetic term in (\ref{fullaction}) takes the form
\bea
{\cal L}_{\rm kin} &=& \frac1{240}\,e\,g^{\mu\nu}\,{D}_{\mu}{\cal M}_{MN}\,{D}_{\nu}{\cal M}^{MN}
 ~=~ -\frac14\,e\,g^{\mu\nu}\,{j}_\mu{}^M \,{j}_{\nu}{\,}_M
\;,  
\label{Lkin}
\eea
in terms of the current ${j}_\mu{}^M$ defined by
\bea
{\cal M}^{KN}{ D}_\mu {\cal M}_{NL} &\equiv& {j}_\mu{}^N\,f_{NL}{}^K
\;,\qquad
\mbox{and satisfying}\quad
{\cal M}_{MN}\,{j}_\mu{}^N = {\eta}_{MN}\,{j}_\mu{}^N
\;.
\label{currJ1}
\eea
All derivatives ${ D}_\mu$ here are covariantized w.r.t.\ generalized internal diffeomorphisms
according to (\ref{covder}), with the matrix ${\cal M}_{MN}$ carrying weight $\lambda=0$.
The second equation in (\ref{currJ1}) can be verified with (\ref{CartanKilling}) and the relation,\footnote{Note the sign, which 
is due to the fact that unlike $\eta_{MN}$ the matrix ${\cal M}_{MN}$ is not a group invariant tensor, but commutes with the 
involution which defines the maximal compact subgroup ${\rm SO}(16)\subset {\rm E}_{8(8)}$\,.
} 
\bea\label{InvStruc}
{\cal M}^{PM} {\cal M}^{QN} \,f_{PQ}{}^K &=& -f^{MN}{}_L\,{\cal M}^{LK}
\;.
\eea

 \paragraph{Chern-Simons term}
The vector fields $A_\mu{}^M$ and $B_{\mu\,M}$  
do not carry propagating degrees of freedom, but describe on-shell
duals to the scalar fields. Consequently their dynamics in (\ref{fullaction})
is not described by a Yang-Mills coupling but rather by
 a topological Chern-Simons term which is explicitly given by
\bea
{\cal L}_{\rm CS} &=&
2\kappa\,\varepsilon^{\mu\nu\rho}\,\Big( 
{F}_{\mu\nu}{}^M B_{\rho}{}_M
-f_{KL}{}^N \partial_\mu A_\nu{}^K \partial_N A_\rho{}^L
-\frac23\,f^N{}_{KL} \partial_M\partial_N A_{\mu}{}^K A_\nu{}^M A_{\rho}{}^L
\nonumber\\
&&{}
\qquad\qquad
-\frac13 \,f_{MKL} f^{KP}{}_Q f^{LR}{}_S\,A_\mu{}^M \partial_P A_\nu{}^Q \partial_R A_\rho{}^S
\Big) 
\;,
\label{CS}
\eea
with coupling constant $\kappa$ that we will determine below.
The structure and covariance of the Chern-Simons term
become more transparent by calculating  its general variation which is given by
\bea
\delta {\cal L}_{\rm CS} &=&
2\kappa\,\varepsilon^{\mu\nu\rho}\,\Big(
{F}_{\mu\nu}{}^M\,\delta B_{\rho\,M}
+({G}_{\mu\nu\,M}-f_{MN}{}^K\partial_K{F}_{\mu\nu}{}^N)\,\delta A_{\rho}{}^M
\Big)
\nonumber\\
&=&
2\kappa\,\varepsilon^{\mu\nu\rho}\,\Big(
{\cal F}_{\mu\nu}{}^M\,\delta B_{\rho\,M}
+({\cal G}_{\mu\nu\,M}-f_{MN}{}^K\partial_K{\cal F}_{\mu\nu}{}^N)\,\delta A_{\rho}{}^M
\Big)
\;.
\label{varyCS}
\eea
Indeed it follows directly with the section constraints (\ref{triv1}) and (\ref{triv2}) that all extra 
two-form contributions proportional to $C_{\mu\nu}$ from (\ref{FGC})
cancel in the second line of (\ref{varyCS}), such that the variation may be expressed entirely in terms of the
covariant quantities. Similarly, one confirms with (\ref{varyCS}) that the Chern-Simons term is 
invariant under the vector shift transformations (\ref{shiftAB}).
With a little more calculation we may furthermore verify invariance of the
Chern-Simons term under generalized internal diffeomorphisms that act as gauge transformations (\ref{gaugeAB})
on the vector fields. 
Specifically, after partial integration, the variation (\ref{gaugeAB}) yields
\bea
\delta {\cal L}_{\rm CS} &=&
2\kappa\varepsilon^{\mu\nu\rho}\,\Lambda^K\,\Big(
{\cal F}_{\mu\nu}{}^M\,(- \partial_M B_{\rho\,K} + f^N{}_{KL}  \partial_M\partial_N A_\rho{}^L)
-{ D}_\rho^{(0)}({\cal G}_{\mu\nu\,K}-f_{KM}{}^N\partial_N{\cal F}_{\mu\nu}{}^M) \Big)
\nonumber\\
&&{}
- 2\kappa\varepsilon^{\mu\nu\rho}\,
\Sigma_M\, { D}_\rho^{(1)} {\cal F}_{\mu\nu}{}^M 
\;.
\label{varCSdiff}
\eea
The vanishing of the r.h.s.\ of this variation corresponds to establishing some generalized 
Bianchi identities for the curvatures (\ref{FGC}). This is most conveniently achieved by
evaluating three covariant derivatives
$\varepsilon^{\mu\nu\rho} {D}_\mu {D}_\nu {D}_\rho V^M$
on a vector $V^M$ of weight 0, from which we deduce the identity
\bea
\varepsilon^{\mu\nu\rho}\,\mathbb{L}_{({\cal F}_{\mu\nu},{\cal G}_{\mu\nu})}\,{D}_\rho V^M
&=&
\varepsilon^{\mu\nu\rho}\,{D}_\mu\left(
\mathbb{L}_{({\cal F}_{\nu\rho},{\cal G}_{\nu\rho})}\, V^M \right)
\;.
\label{DDD}
\eea
Its r.h.s.\ takes the explicit form
\bea
\varepsilon^{\mu\nu\rho}\,{D}_\rho\left(
\mathbb{L}_{({\cal F}_{\mu\nu},{\cal G}_{\mu\nu})}\, V^M \right)
&=&
\varepsilon^{\mu\nu\rho}\,{D}_\rho\left(
{\cal F}_{\mu\nu}{}^N\partial_N V^M - \Big(
{\cal G}_{\mu\nu\,L}-f_{LP}{}^K\partial_K{\cal F}_{\mu\nu}{}^P
\Big)\,f^{LM}{}_N\,V^N
\right)
\;,
\nonumber
\eea
and upon using that
\bea
{D}_\rho \partial_N  V^M  &=&
\partial_N {D}_\rho V^M 
-  f^{LM}{}_{P}\,V^P\,\Big(
\partial_N B_\rho{}\,_L- f_{LQ}{}^K \, \partial_N\partial_K A_\rho{}^Q\Big)
\;,
\eea
for a vector $V^M$ of weight 0, the r.h.s.\ of (\ref{DDD}) may be further rewritten as
\bea
\varepsilon^{\mu\nu\rho}\,{D}_\rho\left(
\mathbb{L}_{({\cal F}_{\mu\nu},{\cal G}_{\mu\nu})}\, V^M \right)
&=&
\varepsilon^{\mu\nu\rho}\,\left(
{\cal F}_{\mu\nu}{}^N\partial_N {D}_\rho V^M - \Big(
{\cal G}_{\mu\nu\,L}-f_{LP}{}^K\partial_K{\cal F}_{\mu\nu}{}^P
\Big)\,f^{LM}{}_N\,{D}_\rho V^N
\right)
\nonumber\\
&&{}\hspace{-0.7cm} 
+\varepsilon^{\mu\nu\rho}\,\left(
{D}_\rho {\cal F}_{\mu\nu}{}^N\partial_N V^M - {D}_\rho \Big(
{\cal G}_{\mu\nu\,L}-f_{LP}{}^K\partial_K{\cal F}_{\mu\nu}{}^P
\Big)\,f^{LM}{}_N\,V^N
\right)
\nonumber\\
&&{}\hspace{-0.7cm} 
-  \varepsilon^{\mu\nu\rho}\, f^{LM}{}_{N}\,V^N\,{\cal F}_{\mu\nu}{}^P\, \Big(
\partial_P B_\rho{}\,_L- f_{LQ}{}^K \, \partial_P\partial_K A_\rho{}^Q\Big)
\;. 
\label{mid2}
\eea
Now the first line in (\ref{mid2}) reproduces the l.h.s.\ of (\ref{DDD}),
such that together we obtain the generalized Bianchi identities
\bea
0&=& \varepsilon^{\mu\nu\rho}\,
{D}^{(1)}_\rho {\cal F}_{\mu\nu}{}^N \otimes \partial_N \;,
\label{genBianchi}\\
0&=& \varepsilon^{\mu\nu\rho}\,\Big(
{D}_\rho^{(0)}({\cal G}_{\mu\nu\,K}-f_{KM}{}^N\partial_N{\cal F}_{\mu\nu}{}^M)
+{\cal F}_{\mu\nu}{}^M\,(\partial_M B_{\rho\,K} - f^N{}_{KL}  \partial_M\partial_N A_\rho{}^L)
 \Big)
\;.\nonumber
\eea
 These are sufficient to show that (\ref{varCSdiff}) vanishes, confirming that the Chern-Simons term
 is invariant under generalized internal diffeomorphisms.
 Let us finally note that a more compact presentation of the Chern-Simons term (\ref{CS}) can be given
 as the boundary contribution of a gauge invariant exact form in four dimension as
\bea 
   S_{\rm CS} &\propto& \int_{\Sigma_4} d^4 x \,\int d^{248} Y\,
   \Big({\cal F}^M\wedge {\cal G}_M-\frac{1}{2}f_{MN}{}^{K} {\cal F}^M\wedge \partial_K {\cal F}^N \Big)
 \;,
    \label{CS4}
\eea
where again all two-form contributions $C_{\mu\nu}$ can be checked to drop out from the action.
Gauge invariance of (\ref{CS4}) follows from the transformation behavior of the field strengths
under gauge transformations
 \bea
  \delta_{\Lambda,\Sigma} \,{\cal F}^{M} 
   &=& \mathbb{L}_{(\Lambda,\Sigma)} \,{\cal F}^{M}\;, 
   \nonumber\\
 \delta_{\Lambda,\Sigma} \big({\cal F}^{M} f_{M}{}^{N}{}_{K} \partial_{N}{\cal F}^{K}\big)
   &=& \mathbb{L}_{(\Lambda,\Sigma)} \big({\cal F}^{M} f_{M}{}^{N}{}_{K} \partial_{N}{\cal F}^{K}\big)
+2\,{\cal F}^{M}\partial_M\partial_N\Lambda^K\, f^{N}{}_{LK}{\cal F}^{L}
 \nonumber\\
 &&{} -2\,{\cal F}^{M}\partial_M\Sigma_N\,{\cal F}^{N}
  \;,
  \nonumber\\
\delta_{\Lambda,\Sigma}\,{\cal G}_{M} &=& \mathbb{L}_{(\Lambda,\Sigma)}\,{\cal G}_{M}
-f^{N}{}_{KL}  {\cal F}^{K}\, \partial_M\partial_N\Lambda^L\, 
  +\partial_M\Sigma_N\,{\cal F}^{N}\;.
  \label{delFFG}
\eea

 \paragraph{Scalar potential}

The last term in the action (\ref{fullaction}) is the scalar potential $V$ which can be given as 
a function of the external metric $g_{\mu\nu}$ and the internal metric ${\cal M}_{MN}$ 
\bea
  V & = &
  -\frac{1}{240}{\cal M}^{MN}\partial_M{\cal M}^{KL}\,\partial_N{\cal M}_{KL}+
  \frac{1}{2}{\cal M}^{MN}\partial_M{\cal M}^{KL}\partial_L{\cal M}_{NK} \label{VinM}\\
  &&{}
  +\frac1{7200}\,f^{NQ}{}_P f^{MS}{}_R\,
  {\cal M}^{PK} \partial_M {\cal M}_{QK} {\cal M}^{RL} \partial_N {\cal M}_{SL} 
  \nonumber\\
  &&{}
  -\frac{1}{2}g^{-1}\partial_Mg\,\partial_N{\cal M}^{MN}-\frac{1}{4}  {\cal M}^{MN}g^{-1}\partial_Mg\,g^{-1}\partial_Ng
  -\frac{1}{4}{\cal M}^{MN}\partial_Mg^{\mu\nu}\partial_N g_{\mu\nu}\;. 
  \nonumber
\eea
The relative coefficients in this potential are determined by $\Lambda^M$ and $\Sigma^M$ gauge invariance
by a computation similar to the one presented for the ${\rm E}_{6(6)}$,  ${\rm E}_{7(7)}$ potentials in
\cite{Hohm:2013vpa,Hohm:2013uia}, that we briefly sketch in the following.
For the calculation it turns out to be convenient to rewrite the potential as
\bea
  V & = &
 \frac{1}{4}{j}_M{}^R{j}_N{}^S
\left({\cal M}^{MN} 
\eta_{RS}
-2{\cal M}^{KL}  
 f_{RL}{}^N f_{SK}{}^M+2\delta_R{}^{N}\delta_{S}{}^{M}
\right)
  \label{VinJ}\\  &&{}
  -\frac{1}{2}g^{-1}\partial_Mg\,{\cal M}^{MN}f_{NK}{}^{P} {j}_{P}{}^{K}
  -\frac{1}{4}  {\cal M}^{MN}g^{-1}\partial_Mg\,g^{-1}\partial_Ng
  -\frac{1}{4}{\cal M}^{MN}\partial_Mg^{\mu\nu}\partial_N g_{\mu\nu}\;,
\nonumber
\eea
in terms of the current ${j}_M{}^P$ defined in analogy to (\ref{currJ1}) as
\bea
{\cal M}^{KP}\partial_M {\cal M}_{PL} &\equiv& {j}_M{}^P\,f_{PL}{}^K
\;.
\label{currJ2}
\eea
A short calculation shows that the
non-covariant variation of the current ${j}_M{}^N$ 
under generalized diffeomorphisms (\ref{DV}) is given by
\bea
\Delta^{\rm nc} {j}_M{}^N &=& 
\left(
{\cal M}^{NK}+\eta^{NK}\right) \,
\partial_M \left(f_{KQ}{}^{P}\partial_P\Lambda^Q - \Sigma_K
\right)
\;,
\eea
where we have used the invariance property (\ref{InvStruc}) of the structure constants. 
It is then straightforward to verify that the non-covariant contributions from the variation
of the various terms in (\ref{VinJ}) precisely cancel.
In particular, we find that under $\Lambda$ transformations the first line of (\ref{VinJ}) transforms
according to
 \bea 
&&
\frac{1}{4}\, \Delta^{\rm nc}_{\Lambda}\Big({j}_M{}^R{j}_N{}^S
\left({\cal M}^{MN} 
\eta_{RS}
-2{\cal M}^{KL}  
 f_{RL}{}^N f_{SK}{}^M+2\delta_R{}^{N}\delta_{S}{}^{M}
\right)\Big) \nonumber\\
 &&\qquad \qquad \qquad =~ 2\,\partial_M\partial_P\Lambda^P\,\partial_N{\cal M}^{MN}
+\partial_M\partial_L\Lambda^N\,\partial_N{\cal M}^{ML}\;,
\eea
whereas the second line of (\ref{VinJ}) transforms into
 \bea
\Delta^{\rm nc}_{\Lambda} \left((\ref{VinJ}), \text{second line}\right)&=&
-3\partial_M\partial_K\Lambda^K\,\partial_P{\cal M}^{PM}
-e^{-1}\partial_Me \,{\cal M}^{MP}\partial_P\partial_R\Lambda^R
\nonumber\\
&&{}
+e^{-1}\partial_M e {\cal M}^{SP}\partial_P\partial_S\Lambda^M
\;.
 \eea
Together, this shows that the scalar potential term $(e\, V)$ in the Lagrangian  is invariant up to total derivatives.

Comparing the expression of (\ref{VinM}) to other results in the literature~\cite{Godazgar:2013rja} shows that the third term 
of (\ref{VinM}) has been missed in previous constructions. Here, this term is essential for $\Lambda^M$ and
$\Sigma_M$ invariance of the scalar potential. Absence of this term is the reason for the observed discrepancy of the 
scalar potential of~\cite{Godazgar:2013rja} with $D=11$ supergravity as we discuss in more detail in the last section. 


\paragraph{External diffeomorphism invariance}


The various terms of the EFT action (\ref{fullaction}) have been determined
by invariance under generalized internal $\Lambda^M$, $\Sigma_M$ diffeomorphisms. 
In contrast, the relative coefficients between the four terms are determined by
invariance of the full action under the remaining gauge symmetries, which are a covariantized 
version of the external $(2+1)$-dimensional diffeomorphisms with parameters $\xi^{\mu}(x,Y)$. 
For $Y$-independent parameter, external diffeomorphism invariance is manifest. 
On the other hand, gauge invariance for general $\xi^{\mu}(x,Y)$ determines all
relative coefficients, as we shall demonstrate in the following.
The computation closely follows the analogous discussion for the ${\rm SL}(2,\mathbb{R})$-covariant
formulation of four-dimensional Einstein gravity~\cite{Hohm:2013jma}.\footnote{We note that here we use 
a field basis for $A$ and $B$ that is related to the ${\rm SL}(2,\mathbb{R})$ treatment of \cite{Hohm:2013jma} 
by a field redefinition.}

Under general external diffeomorphisms, the external and internal metric 
transform in the standard (but covariantized) way
  \be\label{xidiff}
  \delta_{\xi}{\cal M}_{MN} \ = \ \xi^{\mu} {D}_{\mu}{\cal M}_{MN}\;, \qquad
  \delta_{\xi}e_{\mu}{}^{a} \ = \ \xi^{\rho}{D}_{\rho} e_{\mu}{}^{a}+{D}_{\mu}\xi^{\rho} e_{\rho}{}^{a}\;,
 \ee
where we recall that the dreibein is an ${\rm E}_{8(8)}$ scalar-density of weight $\lambda=1$.
The transformation behavior of the gauge vectors is more complicated. 
Inspired by the ${\rm SL}(2,\mathbb{R})$ case \cite{Hohm:2013jma}, for these fields we start from the ansatz
\bea
 \delta^{(0)}_{\xi} A_{\mu}{}^{M} &=& 
 \xi^{\nu}F_{\nu\mu}{}^{M}+{\cal M}^{MN}g_{\mu\nu}\partial_{N}\xi^{\nu} \;,
\nonumber\\
\delta^{(0)}_{\xi} B_{\mu\,M} &=& \xi^{\nu}G_{\nu\mu\,M}
-j_M{}^K
\,g_{\mu\nu} \partial_K\xi^\nu 
+\frac1{4\kappa}\, e\varepsilon_{\mu\nu\lambda}
\,g^{\lambda\rho}\, {D}^\nu\left( g_{\rho\sigma} \partial_M \xi^\sigma \right) 
 \;,
  \label{deltaxiA0}
  \eea
where the non-covariant contributions will be required for particular cancellations in the variation
of the Lagrangian. The full variation of these fields will be determined as we go along. Note that the
form of the variation $\delta^{(0)}_{\xi} B_{\mu\,M}$ is manifestly compatible with the constraints (\ref{secconstr})
which this field is required to satisfy, because in the extra non-covariant terms the external index is carried 
by a derivative.

Let us now compute the variation of the Lagrangian (\ref{fullaction}) under (\ref{xidiff}), (\ref{deltaxiA0}).
To start with, let us work out the general variation of the Lagrangian (\ref{fullaction})
w.r.t.\ to the vector fields which takes the form
\bea
\delta{\cal L} &=& 
\varepsilon^{\mu\nu\rho}\left( {\cal E}_{\nu\rho}^{(A)\,M}\,\delta B_{\mu\,M}
+{\cal E}_{\nu\rho\,M}^{(B)}\,\delta A_{\mu}{}^{M} \right)
\;,
\label{varLAB}
\eea
with
\bea
{\cal E}_{\mu\nu}^{(A)\,M} &\equiv& 
2\,\kappa\,{F}_{\mu\nu}{}^M
+\frac{1}2\, e\,\varepsilon_{\mu\nu\rho}\,  
{j}^\rho{}^M
\;,
\nonumber\\
{\cal E}_{\mu\nu\,M}^{(B)} &\equiv& 
2\,\kappa\,{\cal G}_{\mu\nu\,M} 
-f_{MN}{}^K\partial_K{\cal E}_{\mu\nu}^{(A)\,N}
-\frac{1}{4}\, e\,\varepsilon_{\mu\nu\rho}\left(
j_M{}^K\,j^\rho{}_K +2\,\widehat{J}^\rho{}_{M} \right)
\;,
\label{EAB}
\eea
with the currents $j_\mu{}^N$, $j_M{}^N$ from (\ref{currJ1}) and (\ref{currJ2}), respectively, 
and the current $\widehat{J}^\mu{}_{M}$ describing the
contribution from the covariantized Einstein-Hilbert term, 
\bea
\delta_A {\cal L}_{\rm EH} &\equiv& e \widehat{J}^\mu{}_{M}\,\delta A_\mu{}^M
~=~
-2  e e_{a}{}^\mu e_{b}{}^\nu \left(
\partial_M \omega_{\nu}{}^{ab} 
-  {D}_{\nu} \left(   e^{\rho[a} \partial_M e_{\rho}{}^{b]} \right) \right)
\delta A_\mu{}^M
\;.
\label{Jhat}
\eea
Note that not all components of ${\cal E}_{\mu\nu}^{(A)\,M}$ in (\ref{EAB}) correspond to real equations of motion
of the theory, as the field $B_{\mu\,M}$ is constrained by means of (\ref{secconstr}).

Next we consider the non-covariant variation of the covariantized Einstein-Hilbert term, which 
is given by \cite{Hohm:2013jma}
\bea
  \delta^{(0)}_{\xi}\big(e\widehat{R}\,\big) &=& e F^{\mu\nu N} { D}_{\mu}\left(\partial_N\xi^{\rho} g_{\rho\nu}\right)
  +e{\cal M}^{MN}\,\widehat{J}_\mu{}_{M}\,\partial_{N}\xi^{\mu}
  \;,
 \label{totalEHA}
 \eea  
where the second term comes from the non-covariant transformation (\ref{deltaxiA0})
of the vector field $A_\mu{}^M$  via (\ref{Jhat}). The non-covariant
variation of the Chern-Simons term follows from (\ref{varyCS}) and yields
\bea
\delta^{(0)}_{\xi} {\cal L}_{\rm CS} &=&
-e\,
{ F}^{\mu\nu}{}^M
\, { D}_\mu\left( g_{\nu\sigma} \partial_M \xi^\sigma \right) 
-2\kappa\,\varepsilon^{\mu\nu\rho}\,
{ F}_{\mu\nu}{}^M\,{j}_M{}^K
\,g_{\rho\lambda} \partial_K\xi^\lambda
\nonumber\\
&&{}
-2\kappa\,\varepsilon^{\mu\nu\rho}\,
f_{MN}{}^K\partial_K{ F}_{\mu\nu}{}^N\,{\cal M}^{ML}g_{\rho\sigma}\partial_{L}\xi^{\sigma}
+ 2\kappa\,\varepsilon^{\mu\nu\rho}\,{ G}_{\mu\nu\,M}
{\cal M}^{MN}g_{\rho\sigma}\partial_{N}\xi^{\sigma}
\nonumber\\
&&{}
+\kappa\,\varepsilon^{\mu\nu\rho}\,\partial_K \xi^{\sigma}\,
f_{MN}{}^K{ F}_{\mu\nu}{}^M F_{\rho\sigma}{}^{N}
\;,
\label{varxiCS}
\eea
up to total derivatives. The first term cancels against the contribution from (\ref{totalEHA}).
Let us further rewrite the last term of (\ref{varxiCS}) in terms of (\ref{EAB}) as
\bea
\kappa\,\varepsilon^{\mu\nu\rho}\,\partial_K \xi^{\sigma}\,
f_{MN}{}^K{ F}_{\mu\nu}{}^M F_{\rho\sigma}{}^{N}
&=&
\frac{1}{4\kappa}\,\varepsilon^{\mu\nu\rho}\,\partial_K \xi^{\sigma}\,
f_{MN}{}^K{\cal E}^{(A)\,M}_{\mu\nu}{} {\cal E}^{(A)\,N}_{\rho\sigma}{}
\nonumber\\
&&{}
-\partial_K \xi^{\mu}\,
f_{MN}{}^K\, {j}^{\nu M} \, F_{\mu\nu}{}^N
\nonumber\\
&&{}
-\frac{1}{8\kappa}\,\varepsilon_{\mu\nu\rho}\,
\partial_K \xi^{\mu}\,
f_{MN}{}^K {j}^{\nu M} {j}^{\rho N}
\;.
\label{varxiCS2}
\eea
For the variation of the scalar kinetic term, we start from the variation
\bea
  \delta^{(0)}\big({D}_{\mu}{\cal M}_{MN}\big) & = &
   {\cal L}_\xi \big({ D}_{\mu}{\cal M}_{MN}\big) + \xi^\nu\,[{D}_\mu,{ D}_\nu ] \,{\cal M}_{MN}
   -\mathbb{L}_{( \delta^{(0)}_{\xi} \!A_{\mu}{}^{M}, \delta^{(0)}_{\xi} \! B_{\mu}{}^{M})} {\cal M}_{MN}
   \;,
\eea
which induces the following variation of the kinetic term (\ref{Lkin}):
\bea
\delta^{(0)}_{\xi} {\cal L}_{\rm kin} &=&
\frac{1}{2}\,e\,{\cal M}^{KL} j_K{}^N j_\mu{}_N \partial_L \xi^\nu 
+e\,f^{MK}{}_L\,j^\mu{}_M\,F_{\mu\nu}{}^L\,\partial_K\xi^\nu
+ e\,j_L{}^K \,j_\mu{}^L\,\partial_K \xi^\mu
\nonumber\\
&&{}
-e\,f^{MK}{}_L \,j^\mu{}_M\,\partial_K\left(
{\cal M}^{LN} g_{\mu\nu} \partial_N\xi^\nu \right)
-\frac1{4\kappa}\,\varepsilon^{\mu\nu\rho} \, j_\mu{}^L\,
{ D}_\nu \left(g_{\rho\sigma} \partial_L\xi^\sigma\right)
\;.
\eea
Upon integration by parts, the last term gives rise to
\bea
-\frac1{4\kappa}\,\varepsilon^{\mu\nu\rho} \,
g_{\rho\sigma} \partial_L\xi^\sigma\,{ D}_\mu j_\nu{}^L
&=& 
\frac1{480\,\kappa}\,\varepsilon^{\mu\nu\rho}\,g_{\rho\sigma}\partial_L\xi^\sigma\,f^{KL}{}_M\,{\cal M}_{KN}\,
[{ D}_\mu,{ D}_\nu]\,{\cal M}^{MN}
\nonumber\\
&&{}
+\frac{1}{8\kappa}\,\varepsilon^{\mu\nu\rho}\,g_{\rho\sigma}
\partial_K \xi^{\sigma}\,
f_{MN}{}^K {j}_\mu{}^{M} {j}_\nu{}^{N}
\;,
\eea
and evaluating the commutator of covariant derivatives yields terms that 
precisely cancel the three terms linear in $F_{\mu\nu}{}^M$ and $G_{\mu\nu\,M}$
from (\ref{varxiCS}), provided we choose 
\bea
\kappa &\equiv& \frac14\,,
\eea
for the coupling constant of the CS term.
Putting everything together, for the variation of the first three terms of the Lagrangian (\ref{fullaction})
we find up to total derivatives
\bea
\delta^{(0)}_{\xi}\left(
{\cal L}_{\rm EH}+{\cal L}_{\rm CS}+{\cal L}_{\rm kin}
\right)
&=&
\frac12 \,e\left(
{\cal M}^{KL}   \eta_{RS} 
+2\,\delta_R^K \delta_S^{L}  
-2\, {\cal M}^{MN}\,  f^{K}{}_{MR}f^{L}{}_{NS}
\right)
j_\mu{}^S\,j_{L}{}^{R}\,
\partial_K\xi^\mu 
\nonumber\\
&&{}
+e\,{\cal M}^{MN}\left(
\widehat{J}_\mu{}_{M}\,\partial_{N}\xi^{\mu}-f^{KL}{}_M \,j^\mu{}_K\,\partial_L\left(
g_{\mu\nu} \partial_N\xi^\nu \right)
\right)
\nonumber\\
&&{}
+\varepsilon^{\mu\nu\rho}\,\partial_K \xi^{\sigma}\,
f_{MN}{}^K{\cal E}^{(A)\,M}_{\mu\nu}{} {\cal E}^{(A)\,N}_{\rho\sigma}{}
\;.
\label{deltasum}
\eea
It remains to compare this variation to the non-covariant variation of the scalar potential (\ref{VinM})
under (\ref{xidiff}). Noting that
 \bea
 \label{diffdM}
  \delta_{\xi}(\partial_K {\cal M}_{MN}) &=& \xi^{\mu}{  D}_{\mu}(\partial_K{\cal M}_{MN})
  +\partial_K\xi^{\mu} \,{  D}_{\mu}{\cal M}_{MN}
  \;,\nonumber\\
\delta_{\xi} (\partial_M g_{\mu\nu}) &=&
{\cal L}_\xi (\partial_M g_{\mu\nu}) + (\partial_M\xi^\rho) { D}_\rho g_{\mu\nu}+2\,\partial_M{ D}_{(\mu}\xi^{\rho}\,g_{\nu)\rho}
\;,
\eea
it is straightforward to see from (\ref{VinJ}) that the non-covariant variation of the potential 
due to $\delta^{\rm nc}_{\xi}(\partial_K {\cal M}_{MN})$
precisely cancels the first line of (\ref{deltasum}). Upon further calculation, the remaining contributions from variation
of the potential combine with (\ref{deltasum}) into
\bea
\delta^{(0)}_{\xi} {\cal L} &=&
\Big(e\widehat{J}_\mu{}_{M}
-2e\,{ D}_\mu\left(e^{-1} \partial_M e\right)  -{ D}_\nu\left(e g_{\mu\rho} \partial_M g^{\nu\rho}\right)
+\frac12e\,{ D}_\mu g_{\nu\rho} \partial_M g^{\nu\rho}
\Big)\,
{\cal M}^{MN} \,\partial_N\xi^\mu
\nonumber\\
&&{}
+\varepsilon^{\mu\nu\rho}\,\partial_K \xi^{\sigma}\,
f_{MN}{}^K{\cal E}^{(A)\,M}_{\mu\nu}{} {\cal E}^{(A)\,N}_{\rho\sigma}{}
\;.
\label{deltasum2}
\eea
Using the definite expression (\ref{Jhat}) for $\widehat{J}_\mu{}_{M}$, an explicit calculation shows
that the first line of (\ref{deltasum2}) vanishes identically. We have thus shown that under external
diffeomorphisms (\ref{xidiff}), (\ref{deltaxiA0}), the variation of the Lagrangian (\ref{fullaction}) takes the compact form
\bea
\delta^{(0)}_{\xi} {\cal L}
&=&
\varepsilon^{\mu\nu\rho}\,\partial_K \xi^{\sigma}\,
f_{MN}{}^K{\cal E}^{(A)\,M}_{\mu\nu}{} {\cal E}^{(A)\,N}_{\rho\sigma}{}
\;.
\label{deltasumF}
\eea
Just as in the ${\rm SL}(2,\mathbb{R})$ case we conclude that invariance of the Lagrangian can be achieved
by a further modification of the vector field transformation rules according to \cite{Hohm:2013jma}
\bea
 \delta_{\xi} A_{\mu}{}^{M} &=&  \delta^{(0)}_{\xi} A_{\mu}{}^{M} 
 +  2\, \xi^\nu \, {\cal E}_{\mu\nu}^{(A)\,M} 
\nonumber\\
\delta_{\xi} B_{\mu\,M} &=& \delta^{(0)}_{\xi} B_{\mu\,M}
+  2\, \xi^\nu
\Big({\cal E}_{\mu\nu\,M}^{(B)} 
+f_{MN}{}^K \,\partial_K {\cal E}_{\mu\nu}^{(A)\,N} \Big)
\;.
\label{deltamod}
\eea
It is straightforward to see that the new contributions due to the respective terms in 
$\xi^\nu \, {\cal E}_{\mu\nu}^{(A)\,M}$ and $\xi^\nu \, {\cal E}_{\mu\nu\,M}^{(B)} $  
take the form of an `equations of motion symmetry' and mutually cancel. The last term in (\ref{deltamod})
precisely cancels the variation of (\ref{deltasumF}). Moreover, we note that the new variation $\delta_{\xi} B_{\mu\,M}$
continues to be consistent with the constraints (\ref{secconstr}) that this field is required to satisfy. 

We may summarize the result of this subsection as follows: the action (\ref{fullaction}) is invariant under 
external diffeomorphisms parametrized by $\xi^\mu$ that on the internal and external metric act according to (\ref{xidiff}), 
while their action on the gauge fields follows from combining (\ref{deltaxiA0}) and (\ref{deltamod}), 
\bea
 \delta_{\xi} A_{\mu}{}^{M} &=& 
  e\varepsilon_{\mu\nu\rho}\,\xi^\nu {j}^\rho{}^M
 +{\cal M}^{MN}g_{\mu\nu}\partial_{N}\xi^{\nu} 
 \;,
 \\
\delta_{\xi} B_{\mu\,M} &=& e\varepsilon_{\mu\nu\rho}
\Big(
g^{\rho\lambda}\, { D}^\nu\left( g_{\lambda\sigma} \partial_M \xi^\sigma \right) 
-\frac{1}{2}\, \xi^\nu\,j_M{}^K j^\rho{}_K -  \xi^\nu  \widehat{J}^\rho{}_M
\Big)
-{j}_M{}^K\,g_{\mu\nu} \partial_K\xi^\nu 
 \;.\nonumber
\eea
We have shown that invariance under external diffeomorphisms fixes all the relative
coefficients in (\ref{fullaction}); the action is thus uniquely determined by combining internal and
external generalized diffeomorphism invariance.


\section{Embedding of $D=11$ supergravity}\label{sec4}


In the previous sections, we have constructed the unique ${\rm E}_{8(8)}$-covariant two-derivative
action for the fields (\ref{fieldcontent}), that is invariant under generalized internal and external diffeomorphisms.
It remains to establish its relation to $D=11$ supergravity.
Evaluating the field equations descending from (\ref{fullaction}) 
for an explicit appropriate solution of the section constraints (\ref{secconstr}),
one may recover the full dynamics of $D=11$ supergravity after rearranging the eleven-dimensional fields according to a
$3+8$ Kaluza-Klein split of the coordinates, but retaining the full dependence on all eleven coordinates
as first explored in \cite{Nicolai:1986jk,Koepsell:2000xg}.
We have done this analysis in all detail in the ${\rm E}_{6(6)}$-covariant construction \cite{Hohm:2013vpa}
and reproduced the full and untruncated action of eleven-dimensional supergravity from the ${\rm E}_{6(6)}$ EFT
after various redefinitions and redualizations of fields. 
Here, we keep the discussion brief, sketching the essential steps for the embedding of $D=11$ supergravity
and concentrating on the novel features of the E$_{8(8)}$ case. The complete analysis is left for future work.

The relevant solution of the section condition (\ref{secConstrIntro}) is related to the splitting of coordinates according to the
decomposition of the adjoint representation of  ${\rm E}_{8(8)}$ under its maximal ${\rm GL}(8)$ subgroup:
\bea
{\bf 248} &\longrightarrow&
8_{+3} \oplus 28'_{+2} \oplus 56_{+1} \oplus (1\oplus63)_0 \oplus 56'_{-1} \oplus 28_{-2} \oplus 8'_{-3}
\;,\nonumber\\
\left\{ Y^M \right\}  &\longrightarrow& \{
y^m, y_{mn}, y^{kmn}, y_m{}^n, y_{kmn}, y^{mn}, y_{m} \}
\;,
\label{dec248A}
\eea
with the subscripts referring to the grading w.r.t.\ the ${\rm GL}(1)\subset {\rm GL}(8)$ generator $t_0$\,.
The section constraints (\ref{secconstr}) 
are solved by truncating the coordinate dependence of all fields and gauge parameters to the coordinates in the ${8}_{+3}$:
\bea
\Phi(x^\mu, Y^M) &\longrightarrow&
\Phi(x^\mu, y^m)\;.
\label{solconA}
\eea
In order to see that this truncation provides a solution for the section constraints (\ref{secconstr}), it is sufficient to observe that in the
decomposition of the ${\bf 3875}$ analogous to (\ref{dec248A}), the space of highest grading is an $8_{+5}$, which shows that
\bea
(\mathbb{P}_{3875})_{MN}{}^{mn} &=& 0
\;.
\eea
Accordingly, for the compensating gauge field constrained by (\ref{secconstr}) 
we set all but the associated 8~components $B_{\mu\,m}$ to zero, 
\bea
&&{}
B_{\mu}{}^{m} \rightarrow 0\;,\qquad
B_{\mu\,mn} \rightarrow 0\;,\quad 
B_{\mu}{}^{mnk} \rightarrow 0\;,\quad\nonumber\\
&&{}
B_{\mu}{\,}_m{}^n \rightarrow 0\;,\quad 
B_{\mu}{\,}_{mnk} \rightarrow 0\;,\quad 
B_{\mu}{}^{mn} \rightarrow 0\;.
\label{solconA1}
\eea
In order to recover the fields of $D=11$ supergravity,
we first express the scalar matrix 
${\cal M}_{MN}= ({\cal V}{\cal V}^T)_{MN}$ in terms of a coset-valued vielbein 
${\cal V}\in {\rm E}_{8(8)}/{\rm SO}(16)$, 
parametrized in triangular gauge associated to the grading of (\ref{dec248A}) as \cite{Cremmer:1997ct}
\bea
{\cal V} &\equiv& {\rm exp} \left[\phi\, t_{(0)}\right]\,{\cal V}_8\;
{\rm exp}\left[c_{kmn}\,t_{(+1)}^{kmn}\right]\,{\rm exp}\left[\epsilon^{klmnpqrs}  c_{klmnpq}\, t_{(+2)\,rs}\right]
\,{\rm exp}\left[\varphi_{m}\,t_{(+3)}^{m}\right]
\;.
\label{V248}
\eea
Here, $t_{(0)}$ is the E$_{8(8)}$ generator associated to the GL(1) grading of (\ref{dec248A}), 
${\cal V}_8$ denotes a general element of the ${\rm SL}(8)\subset {\rm GL}(8)$ subgroup,
whereas the $t_{(+n)}$ refer to the E$_{8(8)}$ generators of positive grading in 
(\ref{dec248A}).\footnote{
Explicit expressions for the matrix exponential (\ref{V248}) 
have been worked out in~\cite{Godazgar:2013rja}.}
The scalar fields $c_{mnk}=c_{[mnk]}$ and $c_{mnklpq}=c_{[mnklpq]}$ 
have an obvious origin in the internal components of the $11$-dimensional 3-form and 6-form.
The scalar fields 
on the other hand represent the degree of freedom dual to the Kaluza-Klein vector fields
$A_\mu{}^m$ in the standard decomposition of the eleven-dimensional metric. Hence, formally they carry the 
degrees of the freedom of the dual graviton~\cite{Curtright:1980yk,Hull:2000zn,West:2001as,Hull:2001iu} 
which can be written in more suggestive form by defining 
\bea
 c_{m,n_1\dots n_8} &\equiv& \epsilon_{n_1\dots n_8}\,\varphi_m
\;.
\label{dualg}
\eea

Similarly, the gauge field $A_\mu{}^M$ is split
according to the decomposition (\ref{dec248A}) into
\bea
\left\{ A_\mu{}^M \right\}  &\longrightarrow& \{
A_\mu{}^m, A_\mu{\,}_{mn}, A_\mu{\,}_{kmnpq}, A_\mu{\,}_m{}^n, A_\mu{}^{kmnpq}, A_\mu{}^{mn}, A_\mu{\,}_{m} \}
\;.
\label{breakA}
\eea
Together with the surviving 8 components from (\ref{solconA1}) we count 256 vector fields
which appears to largely exceed the number of fields with possible eleven-dimensional origin.
Rather, from eleven dimensions we expect only the Kaluza-Klein vector fields $A_\mu{}^m$ together with
gauge fields $A_\mu{\,}_{mn}$ and $A_\mu{\,}_{kmnpq}$ from the 3- and the 6-form, respectively.
Fortunately, many of the fields in (\ref{breakA}) do in fact not enter the Lagrangian (\ref{fullaction}). 
They are pure gauge as a consequence of the invariance of the action under the vector shift symmetry (\ref{shiftAB}).
Indeed, closer inspection
of the covariant derivatives (\ref{covder}) and the Chern-Simons couplings (\ref{CS}) shows that out of (\ref{breakA})
only the components $\{A_\mu{}^m, A_\mu{\,}_{mn}, A_\mu{\,}_{kmnpq}\,, A_\mu{\,}_m{}^n\}$ enter the Lagrangian.
More precisely, the covariant derivatives on the scalar fields evaluated in the parametrisation of (\ref{V248}) are of the 
schematic form
\bea
{ D}_\mu c_{kmn} &=& D_\mu c_{kmn} + \partial_{[k} A_{|\mu|\, mn]} \;,\nonumber\\
{ D}_\mu c_{klmnpq} &=& D_\mu c_{klmnpq} + \partial_{[k} A_{|\mu|\,lmnpq]}
+  \partial_{[k} A_{|\mu|\,lm}\,c_{npq]}
 \;,\nonumber\\
 { D}_\mu \varphi_{m} &=& D_\mu \varphi_{m} + \dots 
 + \partial_n A_{\mu\,m}{}^n + B_{\mu\,m}
 \;,
 \label{Dcomp}
\eea
where we have denoted by $D_\mu$ the derivative covariantized
with the Kaluza-Klein vector field $A_\mu{}^m$ w.r.t.\ eight-dimensional internal diffeomorphisms.
The unspecified terms in (\ref{Dcomp}) refer to nonlinear couplings involving the scalar fields
$c_{kmn}$ and $c_{klmnpq}$.
Integrating out the gauge field $B_{\mu\,m}$ thus not only eliminates all the dual graviton components $\varphi_m$ but simultaneously
eliminates all vector fields $A_\mu{\,}_m{}^n$ from the Lagrangian. 
In this process, it is important that the scalar potential (\ref{VinM}) does not depend on the scalar
fields $\varphi_m$. Indeed, invariance of the Lagrangian under the shift $\varphi_m\rightarrow \varphi_m+c_m$
is a direct consequence of the invariance under generalized diffeomorphisms (\ref{DV})
with parameter~$\Sigma_m$\,.
This illustrates once more the role played by the additional covariantly constrained gauge symmetries $\Sigma_{M}$. 
Their presence and associated gauge connection $B_{\mu\,M}$ allows us  
to establish a covariant duality relation involving the degrees of freedom
from the eleven-dimensional metric and subsequently to 
eliminate the dual graviton degrees of freedom $\varphi_m$ from the Lagrangian.

In turn, this procedure of integrating out $B_{\mu\,m}$ induces a Yang-Mills-type coupling for the vector fields $A_\mu{}^m$ 
in a standard mechanism of three-dimensional supergravities~\cite{Nicolai:2003bp}. To see this, note that 
 the first line of the field equations (\ref{EAB}) precisely
relates the Yang-Mills field strength $F_{\mu\nu}{}^m$ to the scalar current as
\bea
F_{\mu\nu}{}^m &=& - e\varepsilon_{\mu\nu\rho}\,j^{\rho\,m}~=~ - e\varepsilon_{\mu\nu\rho}\,
M^{mn}\,{ D}^{\rho}{} \varphi_n + \dots
\;,
\label{Fj1}
\eea
with $M^{mn}\equiv ({\cal V}_8 {\cal V}_8{}^T)^{mn}$\,.

The resulting Lagrangian then only depends on the fields 
\bea
\{g_{\mu\nu}, {\cal V}_8, c_{kmn}, c_{klmnpq}, A_\mu{}^m, A_\mu{\,}_{mn}, A_\mu{\,}_{kmnpq}\}
\;,
\label{fieldsRed}
\eea
corresponding to the various components of the eleven-dimensional metric, 3-form and 6-form.
Its field equations are proper combinations of the eleven-dimensional field equations and the duality 
equation relating the 3-form and the 6-form. As an example, consider the field equations (\ref{EAB}).
With the first line corresponding to (\ref{Fj1}), we observe that the $(_{mn})$-component of the second line gives rise to
\bea
f_{mn,N}{}^K\partial_K{\cal E}_{\mu\nu}^{(A)\,N} &=& 0 \qquad
\Longrightarrow \qquad
\partial_{[k} \left( F_{|\mu\nu|}{\,}_{mn]} + e\,j^{\rho}{}_{mn]} \,\varepsilon_{\mu\nu\rho} \right) ~=~ 0
\;,
\eea
which can be integrated to the duality equation
\bea
F_{\mu\nu}{\,}_{mn} + e\varepsilon_{\mu\nu\rho}\,j^{\rho}{}_{mn} &=& \partial_{[m} B_{|\mu\nu|\,n]}
\;,
\label{Fj2}
\eea
with an undetermined two-form $B_{\mu\nu\,n}$. The latter can be identified with the corresponding 
component of the eleven-dimensional 3-form. Indeed, further derivation $\epsilon^{\mu\nu\rho} \partial_\rho$ of (\ref{Fj2})
shows that it is compatible with the component
\bea
F_{\mu\nu\rho\,m} &=& e\varepsilon_{\mu\nu\rho}\,\epsilon_{m n_1 \dots n_7} F^{n_1 \dots n_7} + \dots
\;,
\label{Fj3}
\eea
of the eleven-dimensional duality equation (3-form $\leftrightarrow$ 6-form) relating the 
field strength of $B_{\mu\nu\,m}$ on the l.h.s.\
to the 7-form field strength $F_{n_1 \dots n_7} =7\,\partial_{[n_1} c_{c_2 \dots c_7]}+\dots$, 
whose internal derivative $\partial_{n_1} F^{n_1 \dots n_7}$
appears as a source in the field equation for $\partial_\mu j^{\mu}{}_{mn}$.
Equations (\ref{Fj2}) and (\ref{Fj3}) can further be used to eliminate all components $c_{klmnpq}$,
$A_{\mu\,kmnpq}$ from the eleven-dimensional 6-form from the equations, and the resulting equations
of motion coincide with those coming from $D=11$ supergravity with its standard field content.

On the level of the action, we get further confirmation from inspecting the scalar potential (\ref{VinM}).
After parametrization (\ref{V248}) of the 248-bein, evaluation of (\ref{solconA}), and truncation of the external metric $g_{\mu\nu}$
to a warped Minkowski$_3$ geometry, the potential reduces to the schematic form
 \be\label{reducedPot}
  V_{\rm trunc} \ \sim \ R(g)+F_{(4)}^2+F_{(7)}^2\;, 
 \ee 
reproducing the contributions from the $D=11$ kinetic terms and Einstein-Hilbert term in the internal directions
in terms of the fields from (\ref{fieldsRed}). This can be directly inferred from the analysis of~\cite{Godazgar:2013rja} 
which obtains for the first line of (\ref{VinM}) the expression (\ref{reducedPot}) up to a term  $F_{\text{dual grav}}^2$ 
resembling a kinetic term for the dual graviton components (\ref{dualg}).
The role of the second line in the full potential (\ref{VinM}) (absent in~\cite{Godazgar:2013rja}) 
is precisely to cancel this unwanted contribution.
Indeed, the form of (\ref{VinJ}) shows that after imposing (\ref{solconA}), the extra term is of the form 
\bea
j_M{}^N j_N{}^M &\rightarrow& 
j_m{}^n j_n{}^m ~=~  M^{ml} M^{nk}\,(\partial_m \varphi_n) (\partial_k \varphi_l) + \dots
~=~ F_{\text{dual grav}}^2
\;.
\eea
Moreover, since the full potential (\ref{VinM}) by construction does not depend on $\varphi_m$,
this confirms the result (\ref{reducedPot}). 

In view of the duality equation (\ref{Fj3}), the last two terms of
the potential (\ref{reducedPot}) both correspond to contributions $F_{klmn}^2$ 
and $F_{\mu\nu\rho\,m}^2$ from the original $D=11$ three-form kinetic term. This shows the necessity of
the $F_{(7)}^2$ term in (\ref{reducedPot}), carrying the contribution of the two-forms $B_{\mu\nu\,m}$ which are not among the 
EFT fields in (\ref{fieldsRed}). The situation is different for the graviton. 
The $D=11$ metric gives rise to the external and internal metric and the Kaluza-Klein vector fields, 
all of which are already encoded in the  E$_{8(8)}$ EFT and show up in (\ref{fieldsRed}). 
Thus, there is no room for the inclusion of a `dual graviton', for this would double the number of metric degrees of freedom. 
Consequently, the match with  $D=11$ supergravity requires that the dual graviton term is absent in (\ref{reducedPot}), as observed here.
We conclude that there is no `dual graviton problem'. 
Summarizing, after rearranging all fields and coordinates of the E$_{8(8)}$ EFT, 
putting the appropriate solution of the section constraint, 
the action may eventually be matched to the one obtained by properly
parametrizing eleven-dimensional supergravity
in the standard $3+8$ Kaluza-Klein split.

Let us finally mention that also IIB supergravity can be embedded into the E$_{8(8)}$ EFT (\ref{fullaction}).
Just as for the ${\rm E}_{6(6)}$ and ${\rm E}_{7(7)}$ EFT \cite{Hohm:2013pua,Hohm:2013vpa,Hohm:2013uia}, there is
another inequivalent solution to the section conditions (\ref{secconstr})
that describes the embedding of the full ten-dimensional IIB 
theory~\cite{Schwarz:1983wa,Howe:1983sra} into the ${\rm E}_{8(8)}$ 
EFT,\footnote{An analogous solution of the ${\rm SL}(5)$ covariant section condition, 
corresponding to some three-dimensional truncation of type IIB,  
was discussed in the truncation of the theory to its potential term \cite{Blair:2013gqa}.
For a more general discussion of section constraints and type IIB solutions see \cite{Strickland-Constable:2013xta}.}
generalizing the situation of type II double field theory \cite{Hohm:2011zr,Hohm:2011dv}. 
For E$_{8(8)}$ the
embedding of the IIB theory goes along similar lines as the $D=11$ embedding described above,
with the relevant
decomposition ${\rm E}_{8(8)}\rightarrow {\rm GL}(7)\times {\rm SL}(2)$ given by
\bea
{\bf 248} &\longrightarrow&
(7,1)_{+4} \oplus (7',2)_{+3}\oplus (35',1)_{+2}\oplus (21,1)_{+1}
\oplus \big( (48,1)\oplus (1,3) \oplus (1,1)\big)_0
\nonumber\\
&&{}
\oplus (21',1)_{-1} \oplus (35,1)_{-2}\oplus (7,2)_{-3} \oplus (7',1)_{-4}
\;.
\eea
The section constraint is then solved by having all fields depend on only the 
coordinates $y^m$ in the $(7,1)_{+4}$ and setting to zero all components of $B_{\mu\,M}$ other than the 
$B_{\mu\,m}$ in the $(7',1)_{-4}$.

\section{Summary and Outlook}\label{sec5}
In this paper we have given the details of the E$_{8(8)}$ exceptional field theory. 
As discussed in detail in the main text, 
the novel feature of this case is that the E$_{8(8)}$ valued generalized metric ${\cal M}_{MN}$
encodes components of the dual graviton but nevertheless allows for a consistent (in particular gauge 
invariant) dynamics thanks to the mechanism of constrained compensator fields 
introduced in \cite{Hohm:2013jma} (that in turn is a duality-covariant extension of the proposal in 
\cite{Boulanger:2008nd}). This mechanism requires the presence of covariantly constrained 
gauge fields, which in the $D=3$ case feature among the gauge vectors  
entering the covariant derivatives. These fields are unconventional, 
but seem to be indispensable for a gauge and duality invariant formulation. 
They are a generic feature of the exceptional field theories, corresponding in each case
to a subset of the $(D-2)$ forms with $D$ denoting the number of external dimensions~\cite{Hohm:2013vpa,Hohm:2013uia}.

Studying the truncations of these theories to the internal sector (i.e.\ neglecting all external coordinate dependence, external metric
and $p$-form fields), it has been a puzzle for a while how ${\rm E}_{8(8)}$ generalized diffeomorphisms might be implemented
as a consistent structure, given that their transformations do not close into an algebra~\cite{Coimbra:2011ky,Berman:2012vc}.
In the full EFT the resolution is remarkably simple. Also in this case there is a gauge-invariant action (\ref{fullaction}) and non-closure of generalized
diffeomorphisms simply indicates an additional symmetry: the covariantly constrained $\Sigma_M$ gauge transformations of (\ref{DV}).
The associated gauge connection $B_{\mu\,M}$ then takes care of the dynamics of the dual graviton degrees of freedom,
just as the analogous $(D-2)$ forms do in higher dimensions.

We have restricted the analysis to the bosonic sector of the theory, where generalized diffeomorphism invariance has
proved sufficient to uniquely determine the action.
We are confident that the extension to include fermions and the construction of a supersymmetric
action is straightforward along the lines of the supersymmetric $D=3$ gauged supergravity~\cite{Nicolai:2000sc}.
The fermions will transform as scalar densities under generalized diffeomorphisms (\ref{DV})
and in the spinor representations of the local `Lorentz group' SO$(1,2)\times {\rm SO}(16)$, 
as in \cite{Marcus:1983hb,Nicolai:1986jk}. For the E$_{7(7)}$ EFT~\cite{Hohm:2013uia} the full supersymmetric
completion has recently been constructed in \cite{GGHNS}.

After completing the detailed construction of exceptional field theory for E$_{d(d)}$, $d=6,7,8$, 
the question arises whether one can go even further, perhaps starting with the affine Kac-Moody group 
E$_{9(9)}$. The pattern of compensating gauge fields in this case would suggest a new set of `covariantly constrained 
scalars' on top of the infinite hierarchy of fields parametrizing the coset space E$_{9(9)}/{\rm K}(E_{9(9)})$. 
We refrain from further speculations.

\section*{Acknowledgments}
The work of O.H. is supported by the 
U.S. Department of Energy (DoE) under the cooperative 
research agreement DE-FG02-05ER41360 and a DFG Heisenberg fellowship. 
We would like to thank Hadi and Mahdi Godazgar, Hermann Nicolai,  and Barton Zwiebach 
for useful comments and discussions.

\section*{Appendix}

\begin{appendix}

\baselineskip 15pt

\section{Closure of E$_{8(8)}$ generalized Lie derivatives}
Before proving closure of the gauge transformations, 
it is convenient to first derive the following Lemma: 
 \be\label{MasterLemma}
 \begin{split}
  f^{R}{}_{UV} f^{UP}{}_{K} f^{VQ}{}_{L}\,\partial_{(P}\otimes \partial_{Q)} \ &\equiv  \ 
  f^{R}{}_{UV} f^{UP}{}_{[K} f^{VQ}{}_{L]}\,\partial_{P}\otimes \partial_{Q} \\
  \ &= \ 
  -\big( 2 \delta_{[K}^{P} f^{QR}{}_{L]}+\eta^{RP}f^{Q}{}_{KL}\big)\partial_{(P}\otimes \partial_{Q)}\;. 
 \end{split}
 \ee 
In order to verify this, we compute by repeated use of the Jacobi identity
 \be
  \begin{split}
   f^{R}{}_{UV} f^{UP}{}_{[K} f^{VQ}{}_{L]} \ &= \ -f^{R}{}_{U}{}^{P} f^{U}{}_{[K|V|} f^{VQ}{}_{L]}
   -f^{R}{}_{U[K} f^{U}{}_{|V|}{}^{P} f^{VQ}{}_{L]} \\
   \ &= \ -\frac{1}{2}f^{R}{}_{U}{}^{P} f^{UQ}{}_{V} f^{V}{}_{KL}
   -f^{R}{}_{U[K}f_{|V|}{}^{UP} f^{V}{}_{L]}{}^{Q}\;. 
  \end{split}
 \ee  
Inserting this form back into (\ref{MasterLemma}) we can apply in each term the lemma (\ref{Lemma}), 
which then yields the right-hand side of (\ref{MasterLemma}). This completes the proof. 
    
Next, we verify closure of the gauge transformations on a vector of weight zero,  
 \be\label{closureApp}
  \big[\,\delta_1,\,\delta_2\,\big]V^M \ = \ \big(\delta_{\Lambda_{12}}+\delta_{\Sigma_{12}}\big)V^M\;, 
 \ee
according to the effective parameters (\ref{EFFimp}). We compute for the left-hand side, 
first including only the $\Lambda$ transformations,  
 \be
 \begin{split}
  \big[\,\delta_1,\,\delta_2\,\big]V^M 
  \ = \ \;&\Lambda_2^K\partial_K\big(\Lambda_1^L\partial_L V^M- f^{M}{}_{NT} 
  f^{TP}{}_{Q}\partial_P\Lambda_1^Q\,V^N\big)\\
  &- f^{M}{}_{NT} f^{TK}{}_{L}\,\partial_K\Lambda_2^L\big(\Lambda_1^P\partial_P V^N-f^{N}{}_{PU} f^{UR}{}_{S}
  \partial_R\Lambda_1^S V^P\big)-(1\leftrightarrow 2)\;. 
 \end{split} 
 \ee 
Some terms cancel directly under the $(1\leftrightarrow 2)$ antisymmetrization, and one finds 
 \be
  \begin{split}
   \big[\,\delta_1,\,\delta_2\,\big]V^M \ = \ \;&\big[\Lambda_2,\Lambda_1\big]^L\partial_L V^M
   -f^{M}{}_{NT} f^{TP}{}_{Q}\,\Lambda_2^K\partial_K\partial_P\Lambda_1^Q\,V^N \\
   & +f^{M}{}_{NT} f^{TK}{}_{L}f^{N}{}_{PU} f^{UR}{}_{S}\,\partial_K\Lambda_2^L\,\partial_R\Lambda_1^S\,V^P
   -(1\leftrightarrow 2)\;. 
  \end{split}
 \ee 
Here, we denoted by $[\;,\;]$ the conventional Lie bracket. It turns out, however, that the 
extra terms in the E-bracket (\ref{EFFimp}), as compared to the Lie bracket, vanish in the transport 
term due to the section constraints, so that the transport term already has  the desired form. 
We find it convenient to work for now with a different but equivalent effective parameter, 
 \be
  \Lambda_{12}^M \ = \  \Lambda_{2}^N\partial_N \Lambda_{1}^M
-7\,  (\mathbb{P}_{3875}){}^{MK}{}_{NL} \,\Lambda_{2}^N \partial_K \Lambda_{1}^L
-\frac1{8}\,\eta^{MK}\eta_{NL}\,\Lambda_{2}^N \partial_K \Lambda_{1}^L -(1\leftrightarrow 2)\;.
 \ee 
Comparing then with the form of the gauge transformation w.r.t.~this $\Lambda_{12}$ 
we read off for 
the remaining terms 
 \be\label{PROOFstep}
  \begin{split}
   \big[\,\delta_1,\,\delta_2\,\big]V^M \ = \ \;& \Lambda_{12}^L\partial_L V^M- f^{M}{}_{NT} 
  f^{TP}{}_{Q}\partial_P\Lambda_{12}^Q\,V^N \\
  &\hspace{-0.25cm}+f^{M}{}_{NT} f^{TP}{}_{Q}\,\partial_P\Lambda_2^L\,\partial_L\Lambda_1^Q\, V^N
  -7\,f^{M}{}_{NT} f^{TP}{}_{Q}\,\mathbb{P}^{QK}{}_{RS}\,\partial_P\big(\Lambda_2^R\partial_K\Lambda_1^S\big)V^N\\
   &\hspace{-0.25cm}
    +f^{M}{}_{NT} f^{TK}{}_{L}f^{N}{}_{PU} f^{UR}{}_{S}\,\partial_K\Lambda_2^L\,\partial_R\Lambda_1^S\,V^P
   -(1\leftrightarrow 2)\;, 
  \end{split}
 \ee 
where here and in the following we omit the representation label on the ${\bf 3875}$ projector $\mathbb{P}$, 
as it can always be distinguished from its index structure.  The terms in the first line are the ones desired for 
closure, while the terms in the second and third line are extra. We next have to show that these are zero or else 
can be brought to the form of $\Sigma_M$ gauge transformations. 

We investigate terms with $\partial\Lambda\partial\Lambda$ and 
$\Lambda\partial\partial\Lambda$ separately. The latter originate from 
the second term in the second line. Inserting the projector (\ref{3875proj}) we compute 
 \be
 \begin{split}
  -7\,&f^{M}{}_{NT} f^{TP}{}_{Q}\,\mathbb{P}^{QK}{}_{RS}\,\Lambda_2^R \partial_P\partial_K\Lambda_1^S \\
  & \ = \ -\frac{1}{2}\,f^{M}{}_{NT} f^{TP}{}_{Q}\Big[2\,\delta_{R}^{(Q}\delta_S^{K)}
  -f^{U}{}_{R}{}^{(Q} f_{US}{}^{K)} \Big]\Lambda_2^R\partial_P\partial_K\Lambda_1^S \;. 
 \end{split}
 \ee 
Writing this out yields four terms, two with $ff$ and two with $ffff$. Using the lemma (\ref{MasterLemma}) 
we can then reduce the $ffff$ terms to $ff$ terms. After some algebra, one finds that all $ff$ terms cancel, 
proving that the $\Lambda\partial\partial\Lambda$ structures in (\ref{PROOFstep}) actually drop out. 
Next, we turn to the $\partial\Lambda\partial\Lambda$ structures. 
The strategy here is to implement the antisymmetry in $(1\leftrightarrow 2)$ by decomposing 
the terms into structures of the form $\partial_{(P}\Lambda_2^{[R}\partial_{K)}\Lambda_1^{S]}$ and 
$\partial_{[P}\Lambda_2^{(R}\partial_{K]}\Lambda_1^{S)}$. In the former, $ffff$ terms can 
then be reduced to $ff$ terms by means of (\ref{MasterLemma}). 
After some algebra, one then finds for the terms in the second line of (\ref{PROOFstep}) 
 \be
 \begin{split}
   -7\,&f^{M}{}_{NT} f^{TP}{}_{Q}\,\mathbb{P}^{QK}{}_{RS}\,\partial_P\Lambda_2^R\,\partial_K\Lambda_1^S
   -(1\leftrightarrow 2)\\
   & \ = \ 
  -2\,f^{M}{}_{NT} f^{TP}{}_{Q}\,\partial_{[P}\Lambda_2^{(Q}\,\partial_{K]}\Lambda_1^{K)}
  -f^{M}{}_{NT} f_{S}{}^{UQ} f_{U}{}^{K}{}_{R} f_{Q}{}^{PT}\,\partial_{[P}\Lambda_2^{(R}\,\partial_{K]}\Lambda_1^{S)}\;. 
 \end{split} 
 \ee  
Combing with the first term in the second line of (\ref{PROOFstep})  one obtains 
 \be\label{interStep1}  
  \begin{split}
   &f^{M}{}_{NT} f^{TP}{}_{Q}\,\partial_P\Lambda_2^L\,\partial_L\Lambda_1^Q
   -7\,f^{M}{}_{NT} f^{TP}{}_{Q}\,\mathbb{P}^{QK}{}_{RS}\,\partial_P\Lambda_2^R\,\partial_K\Lambda_1^S
   -(1\leftrightarrow 2)\\
   &\quad \ = \ 2\,f^{M}{}_{NT} f^{TP}{}_{Q}\,\partial_{(P}\Lambda_2^{[K}\,\partial_{K)}\Lambda_1^{Q]}
  -f^{M}{}_{NT} f_{S}{}^{UQ} f_{U}{}^{K}{}_{R} f_{Q}{}^{PT}\,\partial_{[P}\Lambda_2^{(R}\,\partial_{K]}\Lambda_1^{S)}\;. 
 \end{split} 
 \ee  
Next, we have to simplify the terms in the third line of   (\ref{PROOFstep}).
We first note that the antisymmetrization in $(1\leftrightarrow 2)$ imposes an antisymmetrization 
of the $T,U$ indices in $f^{TK}{}_{L} f^{UR}{}_{S}$. This structure can thus be written as 
 \be\label{A11}
   2\,f^{M}{}_{N[T} f^{TK}{}_{|L}f^{N}{}_{P|U]} f^{UR}{}_{S}\,\partial_K\Lambda_2^L\,\partial_R\Lambda_1^S
   \ = \ -f^{M}{}_{NP} f^{N}{}_{UT} f^{TK}{}_{L} f^{UR}{}_{S}\,\partial_K\Lambda_2^L\,\partial_{R}\Lambda_1^S\;, 
 \ee  
where we used the Jacobi identity for the contraction of the first and third structure constant. 
In this form the antisymmetry in  $(1\leftrightarrow 2)$ is manifest. Next, we can decompose the 
index pair in $\partial_K\Lambda_2 \,\partial_{R}\Lambda_1$ into its symmetric and antisymmetric part. 
Applying then for the symmetric part the lemma (\ref{MasterLemma}), one finds after some straightforward 
algebra that these terms equal 
 \be\label{interStep2} 
 \begin{split}
  {\rm (\ref{A11})} 
  \ = \ \,&f^{M}{}_{NP} f^{N}{}_{TU} f^{TK}{}_{L} f^{UR}{}_{S}\,\partial_{[K}\Lambda_2^L\,\partial_{R]}\Lambda_1^S
  -f^{M}{}_{NP} f^{RN}{}_{S}\,\partial_{(K}\Lambda_2^K\,\partial_{R)}\Lambda_1^S\\
  &+f^{M}{}_{NP} f^{RN}{}_{L}\,\partial_{(K}\Lambda_2^L\,\partial_{R)}\Lambda_1^K
  -f^{MK}{}_{P} f^{R}{}_{LS}\,\partial_{(K}\Lambda_2^L\,\partial_{R)}\Lambda_1^S\;. 
 \end{split} 
 \ee 
Combining now  
(\ref{interStep1}) and (\ref{interStep2}) in the gauge algebra (\ref{PROOFstep}) one finds that the $ffff$ terms, 
after relabeling of indices, 
combine into 
 \be
  -f^{M}{}_{NT}f^{UK}{}_{R}\big(f_{SU}{}^{Q} f_{Q}{}^{PT}-f_{S}{}^{PQ} f_{QU}{}^{T}\big)
  \partial_{[P}\Lambda_2^{(R}\partial_{K]}\Lambda_1^{S)} V^N\;.
 \ee 
The terms in parenthesis can now be combined by the Jacobi identity to give the structure 
$f_{S}{}^{TQ} f_{QU}{}^{P}$, after which it follows with the Lemma (\ref{Lemma}) that this term 
vanishes. 
The $ff$ terms combine to give the following final result 
 \be
   \big[\,\delta_1,\,\delta_2\,\big]V^M \ = \  \delta_{\Lambda_{12}}V^M 
   -f^{MK}{}_{P}\big(f^{R}{}_{LS}\,\partial_{(K}\Lambda_2^L\,
   \partial_{R)}\Lambda_1^S\big) V^P\;. 
 \ee
As required, the extra term can be interpreted as a $\Sigma_M$ gauge transformation, so that we established in 
total 
 \be
  \big[\,\delta_1,\,\delta_2\,\big]V^M \ = \  \delta_{\Lambda_{12}}V^M+\delta_{\Sigma_{12}}V^M\;, \qquad
  \Sigma_{12 M} \ = \ - f^{N}{}_{PQ}\,\partial_{(M}\Lambda_2^P\,\partial_{N)}\Lambda_1^Q\;. 
 \ee 
Note that the  $\Sigma_M$ gauge parameter is manifestly covariantly constrained in that 
its free index is always carried by a derivative. 
This completes the proof of closure. 
Finally, we may redefine these gauge parameters by trivial parameters of the form (\ref{specialTRIV}), 
 \be
  \chi^K \ = \ \frac{1}{4}\,f^{K}{}_{PQ} \Lambda_2^P\Lambda_1^Q\;.
 \ee 
This brings the gauge algebra into the equivalent form (\ref{EFFimp}) that we
used in the main text. 
We finally note that the closure of $\Sigma$ and $\Lambda$ transformations as indicated in 
(\ref{EFFimp}) follows by a straightforward computation that uses the Lemma (\ref{Lemma}) 
for the constrained parameters $\Sigma_M$. This concludes our proof of closure of the 
E$_{8(8)}$ generalized Lie derivatives.

\end{appendix}



\providecommand{\href}[2]{#2}\begingroup\raggedright\endgroup


\end{document}